\documentclass[twocolumn,superscriptaddress,aps,prx,longbibliography,floatfix]{revtex4-2}

\usepackage{amsmath,amssymb,amsfonts}
\usepackage{graphicx}
\usepackage{xcolor}
\usepackage{hyperref}
\usepackage{booktabs}
\usepackage{braket}
\usepackage{tikz}
\usetikzlibrary{arrows.meta,positioning,calc}

\definecolor{metablue}{RGB}{0,141,169}
\hypersetup{colorlinks=true, linkcolor=metablue, citecolor=metablue, urlcolor=metablue}

\newcommand{\ellzero}{\ell_{0}}
\newcommand{\ellB}{\ell_{B}}
\newcommand{\omegac}{\omega_{c}}

\begin{document}

\title{Twisted light generates robust many-body states for practical quantum computing}

\author{F.~J.~Rodr\'{\i}guez}
\affiliation{Departamento de F\'{\i}sica, Universidad de Los Andes, Bogot\'a D.C., Colombia}

\author{L.~Quiroga}
\affiliation{Departamento de F\'{\i}sica, Universidad de Los Andes, Bogot\'a D.C., Colombia}

\author{N.~F.~Johnson}
\email{neiljohnson@gwu.edu}
\affiliation{Physics Department, George Washington University, Washington, DC 20052, USA}

\date{\today}

\begin{abstract}
Twisted light carries orbital angular momentum (OAM) and can drive excitations of confined, interacting electrons that are dark to uniform dipolar probes. Here we show how this ``beyond-Kohn's-Theorem'' optical channel can become a concrete control primitive for quantum computing. Correlation sectors in few-electron quantum dots -- characterized by the relative angular momentum quantum number -- form a tunable ladder of many-body states that are robust in the limited sense of symmetry-protected selection rules and persistent chiral spectroscopic fingerprints; full topological gap protection requires three or more electrons. A twisted-light pulse with prescribed OAM index and polarization provides fast optical write, read, and scalable addressing of these sectors via the selection rule $\Delta|m|=\pm(l+\sigma)$. In the analytically solvable Calogero ($1/r^2$) interaction limit, both the energy spectrum and the twisted-light matrix elements are closed-form functions of the interaction strength, allowing gate parameters (Rabi frequency, qubit frequency, anharmonicity, and leakage rates) to be written down explicitly. We map these results onto a universal single-qubit gate set, propose a concrete two-qubit entangling mechanism via state-dependent Coulomb coupling between adjacent dots, and identify the dominant decoherence channel (quadrupolar charge noise). A semi-analytic $N=3$ extension using the $1/N$ expansion provides a design-level scaffold for the topological roadmap, including quasihole sector addressing. The central operational message is that twisted light enables WRITE (pulse-create a correlation sector), READ (spectroscopically diagnose correlations), and SCALE (optical addressing via spatial light modulator) in a unified photonic control layer. Throughout, screened and Coulomb interactions preserve the same qualitative chiral fingerprints established in the solvable limit.
\end{abstract}

\maketitle

\section{Introduction}

Ultrafast chiral spectroscopy is often formulated through polarization and dipole selection rules.
Twisted light adds chirality via orbital angular momentum (OAM) and can address internal (relative-motion) excitations that remain dark to spatially uniform dipolar driving.\cite{Allen1992,Afanasev2014,Quinteiro2009}
In magnetized quantum dots, these internal excitations are precisely where strong correlations live: they organize the ``magic-number'' angular-momentum spectrum and (for larger particle number $N$) they connect continuously to fractional quantum Hall effect (FQHE) states.\cite{Reimann2002,Johnson1992,Laughlin1983}

The motivation for quantum computing is twofold.

\textit{First}, topological many-body states offer a route to hardware-level robustness: a gap and symmetry structure can suppress certain local error channels.\cite{NayakReview,KitaevFault}
Even before true topological order sets in (which requires $N\ge 3$), few-electron ``Laughlin-like'' correlation sectors provide sharply distinct wavefunctions with different correlation holes and different spatial structure.

\textit{Second}, twisted light provides an \emph{optical} control layer: the same OAM selection rules that make certain transitions visible also make them \emph{addressable}.
This suggests a ``photonic control bus'' concept in which a spatial light modulator (SLM) steers and sets the beam's OAM so that (i)~beam position selects the dot, while (ii)~vorticity selects the transition.
This optical addressing can, in principle, scale without dense on-chip wiring.

The underlying physics channel is established in our recent work on twisted-light chiral spectroscopy of interacting electrons in nanostructures under magnetic field.\cite{Rodriguez2025}
In that setting, twisted light breaks the (generalized) Kohn theorem by coupling to the \emph{relative} coordinate, opening interaction-enabled ``dark'' lines that are absent for uniform dipole probes.\cite{Rodriguez2025}
Here we recast those exact selection rules and matrix elements in the language of quantum information: they define a natural qubit encoding, a universal single-qubit gate set, and suggest a concrete path toward entangling two-qubit gates.

\section{Two-electron quantum dot under magnetic field: why $N=2$ matters}

\begin{figure}[t]
\centering
\includegraphics[width=\columnwidth]{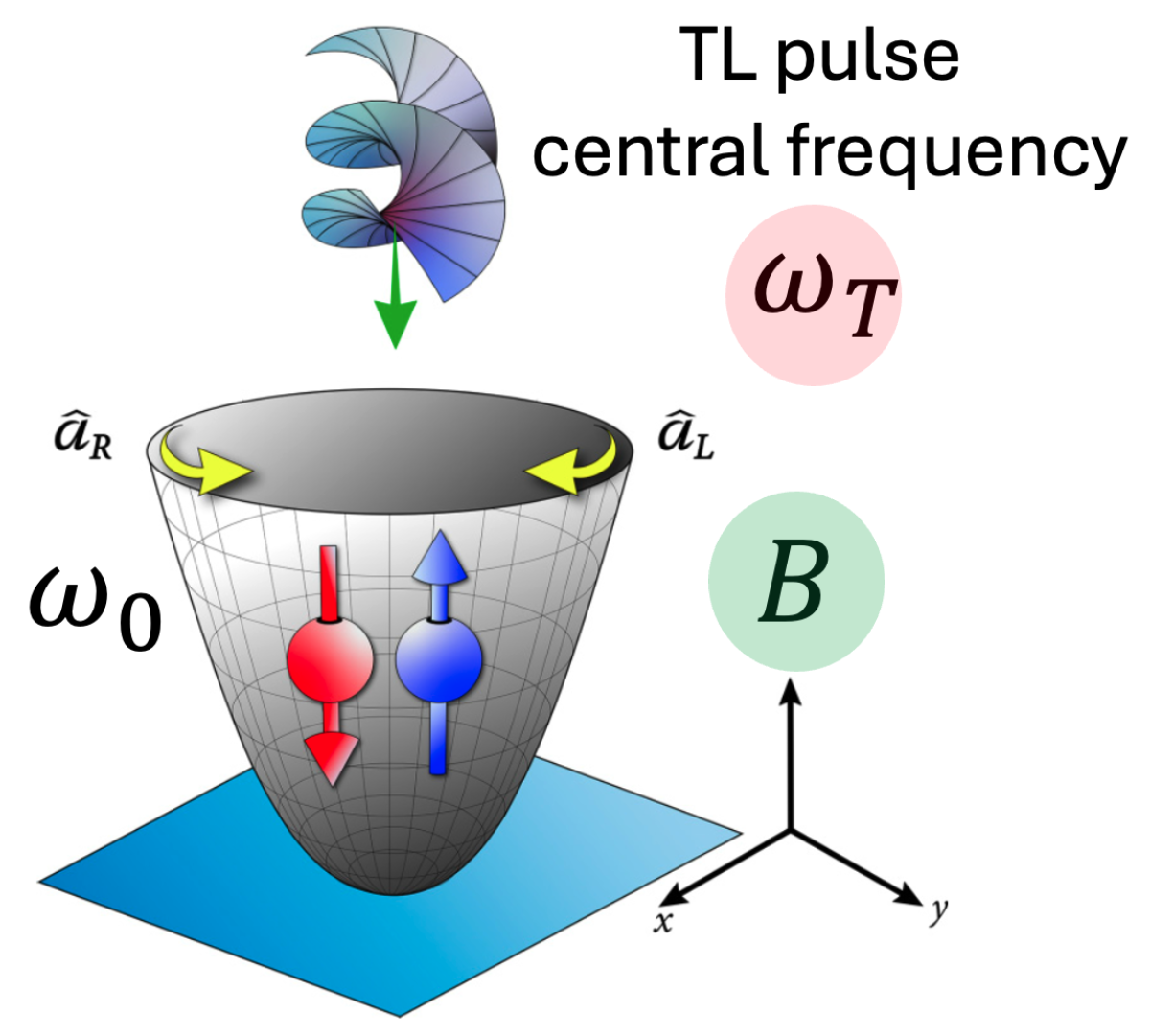}
\vspace{-0.3em}
\caption{Schematic representation of the system. A twisted-light (TL) pulse is incident on two interacting charged electrons confined in a two-dimensional harmonic quantum dot, under a perpendicular magnetic field $B$. The chiral operators $\hat{a}_R$ and $\hat{a}_L$ destroy right- and left-circular orbital quanta of the relative motion, respectively.}
\label{fig:system_schematic}
\vspace{-0.8em}
\end{figure}

\subsection{Center-of-mass vs.\ relative motion and Kohn Theorem protection}

Two electrons of effective mass $m^\ast$ in a 2D parabolic confinement (frequency $\omega_0$) under perpendicular magnetic field $B$ are described by a Hamiltonian that separates exactly into center-of-mass (CM) and relative parts.
Defining complex coordinates $z_j=x_j+iy_j$, and
\begin{equation}
Z=\frac{z_1+z_2}{2},\qquad z=z_1-z_2,
\end{equation}
the non-interacting problem becomes two independent Fock--Darwin oscillators with frequencies
\begin{equation}
\omega = \sqrt{\omega_0^2+\omegac^2/4},\qquad \omegac=\frac{eB}{m^\ast},
\end{equation}
and (in ladder-operator form) right/left circular modes with $\omega_R=\omega+\omegac/2$ and $\omega_L=\omega-\omegac/2$.

The \textbf{generalized Kohn theorem} states that a spatially uniform dipolar probe ($l=0$) excites only the CM degree of freedom in a parabolic dot, independent of interactions.\cite{Reimann2002}
Hence, dipole absorption cannot diagnose the internal correlated spectrum.

Twisted light changes this: because its field has spatial structure (nonzero OAM index $l$), it can carry multipole components that couple directly to the \emph{relative} coordinate.
The resulting absorption therefore fingerprints correlations rather than just the CM mode.\cite{Rodriguez2025}

\subsection{Correlation sectors as ``Laughlin-like'' states at finite N}

In the lowest Landau level (LLL) language, the two-electron antisymmetric polynomial is $(z_1-z_2)$, and the Laughlin-like family is
\begin{equation}
\Psi_{q}^{(N=2)}(z_1,z_2)= (z_1-z_2)^{q}\,
e^{-(|z_1|^2+|z_2|^2)/(4\ellB^2)},
\qquad q\ \text{odd}.
\label{eq:N2_laughlin_family}
\end{equation}
For $N=2$, calling $q=3$ a ``$\nu=1/3$ Laughlin state'' is algebraically correct but physically thin: it has the correct correlation hole but not the topological features (anyon statistics, edge theory, degeneracy) that require $N\ge 3$.\cite{Laughlin1983}
Nevertheless, the \emph{ladder} $q=1\to 3\to 5\to\cdots$ is a sharp, exactly defined set of correlation sectors.
This is precisely the structure that twisted light can address selectively.
\section{Twisted light coupling and selection rules}

A twisted-light mode is characterized by OAM index $l$ and helicity $\sigma=\pm 1$ (circular polarization).
The \emph{total} angular momentum transferred per absorbed photon is $s=l+\sigma$.
For the parabolic-dot geometry, the internal (relative-motion) coupling takes a compact ladder-operator form.\cite{Rodriguez2025}
For example, for the quadrupole-like case ($l=1$ with $\sigma=+1$, hence $s=2$) one obtains
\begin{equation}
\begin{split}
&{\hat H}^{(l=1)}_{2e} = \hbar \omega_R\hat{a}_{R}^{\dagger}\hat{a}_{R} + \hbar \omega_L\hat{a}_{L}^{\dagger}\hat{a}_{L} + V_{e-e}(|\hat{\vec{r}}|) \\
&+i{\cal E}_{0}f(t) \left[ e^{-i\omega_T t}\left(\hat{a}_{R}^{\dagger}+\hat{a}_{L}\right)^2 - e^{i\omega_T t}\left(\hat{a}_{R}+\hat{a}_{L}^{\dagger}\right)^2 \right],
\end{split}
\label{eq:TL_Hamiltonian_l1}
\end{equation}
with a coupling amplitude ${\cal E}_{0}$ (an energy scale) set by the beam amplitude and waist, and $f(t)$ the pulse envelope. This dimensionless light-matter coupling strength is ${\cal E}_{0}= \frac{q_eA_0k_r\ell_0^2\omega_T}{2}$ using the notation of Ref.  
\cite{Rodriguez2025} to which we refer for definition of the parameters.  
The key point is the selection rule:
\begin{equation}
\Delta |m| = \pm s,\qquad s=l+\sigma,
\label{eq:selection_rule}
\end{equation}
so that, for $l=1,\sigma=+1$, twisted light drives $\Delta|m|=\pm 2$ transitions between odd-$|m|$ correlation sectors.
In the left-chiral convention appropriate to the LLL branch for $B>0$, the physical states have $m\le 0$ and we use $|m|=-m$ as a nonnegative sector label; then absorption that \emph{increases} $|m|$ corresponds to $m\rightarrow m-s$.

\begin{figure}[tbh]
\centering
\includegraphics[width=\columnwidth]{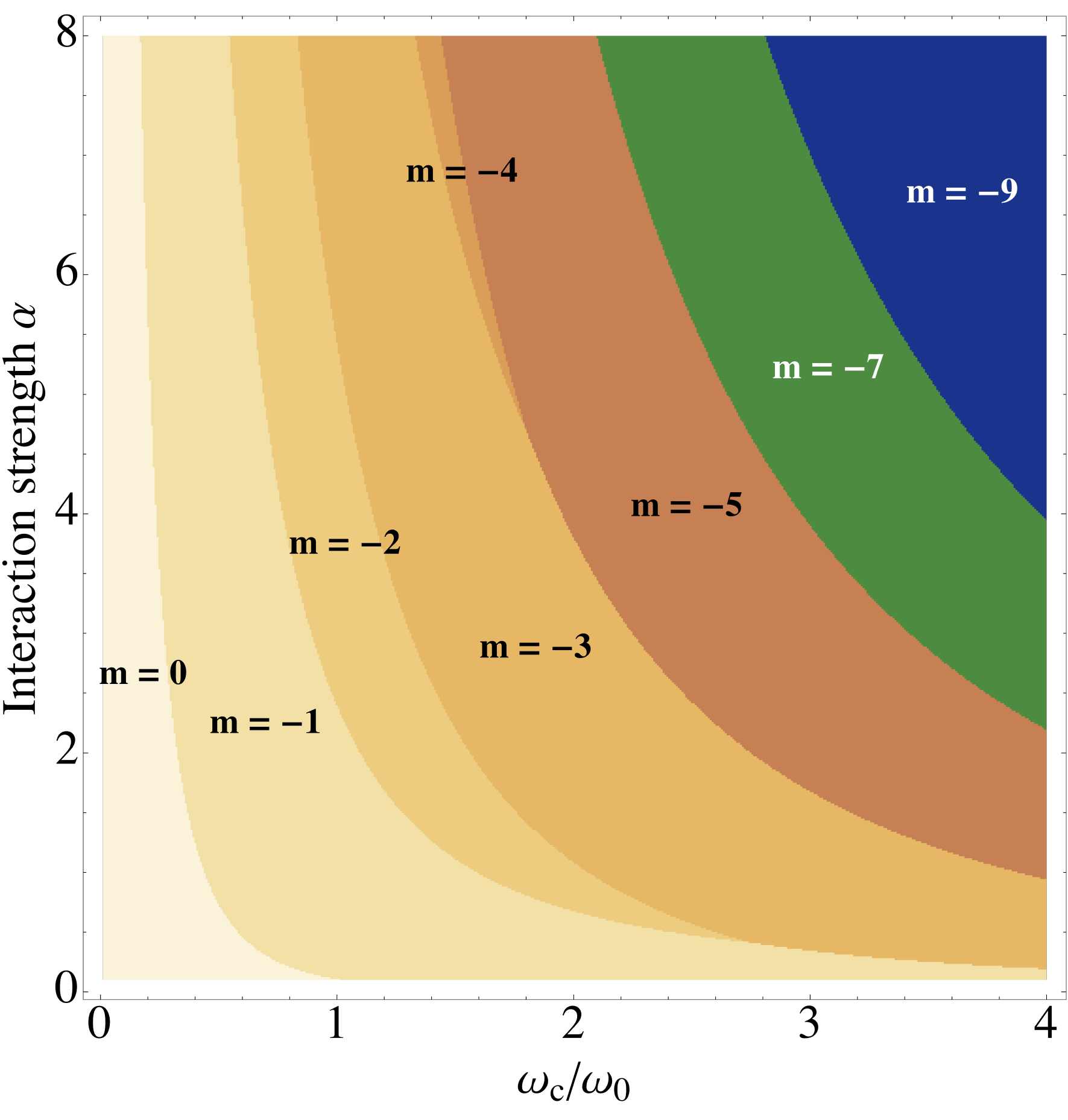}
\caption{Ground-state correlation-sector map. Heat map of the relative-motion ground state in the two-electron parabolic quantum dot as a function of the reduced magnetic field $\omega_c/\omega_0$ and the dimensionless interaction strength $\alpha$. Each colored region identifies the angular-momentum (correlation) sector corresponding to the lowest relative energy (obtained in closed form in the $1/r^2$ interaction limit). The boundaries between regions indicate level crossings where the ground state switches between correlation sectors. These correlation sectors form the ladder of internal states that twisted light addresses selectively via the OAM/helicity selection rule discussed in the main text.}
\label{fig:mGSheatmap}
\vspace{-0.8em}
\end{figure}

With this convention in place, the correlated ground-state structure can be summarized in a two-parameter color map spanning the reduced magnetic field $\omega_c/\omega_0$ and the dimensionless interaction strength $\alpha$ (Fig.~\ref{fig:mGSheatmap}). The colored regions indicate which angular-momentum (correlation) sector $m$ hosts the lowest relative energy $E_{\mathrm{rel}}(n=0,m)$ at each point in parameter space, while the boundaries between regions mark level crossings where the ground state switches between correlation sectors. These boundaries are defined by degeneracy conditions $E_{\mathrm{rel}}(0,m)=E_{\mathrm{rel}}(0,m')$. In the analytically tractable $1/r^2$ limit, the $m$-dependence enters through $\alpha_m=\sqrt{m^2+\alpha}$, together with the confinement renormalisation factor $\sqrt{1+(\omega_c/\omega_0)^2}$ in our dimensionless units. Neglecting the comparatively small Zeeman correction, the crossing condition between two sectors $m$ and $m'$ can be written in the compact form
\begin{equation}
(m-m')\,\frac{\omega_c}{\omega_0}
\;+\;
\sqrt{1+\Big(\frac{\omega_c}{\omega_0}\Big)^2}\,
\big(\alpha_m-\alpha_{m'}\big)
\;=\;0,
\end{equation}
which naturally generates the hyperbola-like curves observed in the map by combining a term linear in $\omega_c/\omega_0$ with a $\sqrt{1+(\omega_c/\omega_0)^2}$ dependence. This representation is particularly useful in the present context because, in a parabolic dot, Kohn’s theorem implies that a spatially uniform dipolar drive couples predominantly to the centre-of-mass degree of freedom and is therefore largely insensitive to the internal, interaction-driven spectrum. Twisted light circumvents this restriction by providing a structured optical field that can couple directly to the relative coordinate, opening a “beyond-Kohn” channel into precisely those internal correlation sectors captured by the map. Moreover, these correlation sectors form the ladder of internal states that twisted light addresses selectively through the OAM/helicity selection rule $\Delta|m|=\pm(l+\sigma)$ discussed above. The heat map therefore identifies the field–interaction windows in which a chosen $|m|$ ladder is energetically preferred and where state-selective twisted-light control is expected to be most clearly defined.

\section{Exactly solvable Calogero limit: closed-form spectrum and matrix elements}

\subsection{Why the Calogero interaction is useful}

A generic screened interaction can be modeled as $V(r)\propto 1/r^\eta$ with $1\le \eta\le 2$.\cite{Rodriguez2025}
The inverse-square limit $\eta=2$ is special: it yields the Calogero interaction
\begin{equation}
V_{\mathrm{Cal}}(r)=\frac{\hbar^2}{2\mu}\,\frac{\alpha}{r^2},
\qquad \mu=m^\ast/2,
\label{eq:calogero_potential}
\end{equation}
for which the relative-coordinate eigenstates are analytic.\cite{Johnson1995,GQR1996}
This solvable limit provides a symmetry ``skeleton'' for the spectroscopy, and the qualitative chiral fingerprints persist for Coulomb-like interactions.\cite{Rodriguez2025}

\subsection{Exact eigenstates and energies}
Separating variables in the relative coordinate as $\psi(r,\phi)=R(r)e^{im\phi}$ reduces the two-dimensional Schr\"odinger problem with $V_{\mathrm{Cal}}(r)\propto \alpha/r^{2}$ to an effective radial equation. 
Regularity at the origin fixes the short-distance scaling $R(r)\sim r^{\alpha_m}$ with $\alpha_m=\sqrt{m^{2}+\alpha}$, while harmonic confinement in a magnetic field enforces a Gaussian envelope $\exp[-r^{2}/(8\ell_{0}^{2})]$ in the lowest radial sector. 
This standard Calogero/Fock--Darwin construction yields the normalised ground state in Eq.~(7) (and, for $n>0$, associated-Laguerre excitations), together with the spectrum in Eq.~(8); see Refs.~(\cite{Johnson1995,GQR1996}) for full derivations.
For the relative coordinate $(r,\phi)$, the Calogero ground state at fixed angular momentum $m$ (radial quantum number $n=0$) is\cite{Johnson1995,GQR1996}
\begin{equation}
\begin{aligned}
\psi^{(\alpha)}_{0,m}(r,\phi)
&=
\frac{e^{im\phi}}{2\sqrt{\pi}\,\ellzero\,\sqrt{\Gamma(\alpha_m+1)}}\,
\left(\frac{r}{2\ellzero}\right)^{\alpha_m}
e^{-r^2/(8\ellzero^2)},\\
\alpha_m&=\sqrt{m^2+\alpha}.
\end{aligned}
\label{eq:calogero_state}
\end{equation}
Here $\ellzero$ is the single-particle Fock--Darwin length, $\ellzero=\sqrt{\hbar/(2m^\ast\omega)}$ with $\omega=\sqrt{\omega_0^2+\omega_c^2/4}$, so that a product of single-particle Gaussians $e^{-\sum_i |z_i|^2/(4\ellzero^2)}$ produces the relative Gaussian $e^{-r^2/(8\ellzero^2)}$.
The corresponding relative energy spectrum is
\begin{equation}
E^{\mathrm{rel}}_{n,m}=\hbar\omega\,(2n+1+\alpha_m)+\frac{m\hbar\omegac}{2},
\label{eq:calogero_energy}
\end{equation}
with $\omega=\sqrt{\omega_0^2+\omegac^2/4}$.
For the physically relevant left-chiral branch at $B>0$ we take $m\le 0$ and label sectors by $|m|=-m$; since $\alpha_m$ depends on $m^2$, only the linear cyclotron term distinguishes the two chiralities.

These closed-form expressions are rare in correlated quantum systems.
They imply that energy splittings between correlation sectors (e.g.\ $|m|=3$ vs.\ $|m|=5$) are analytic functions of the interaction parameter $\alpha$.
\begin{figure}[t]
\centering
\includegraphics[width=\columnwidth]{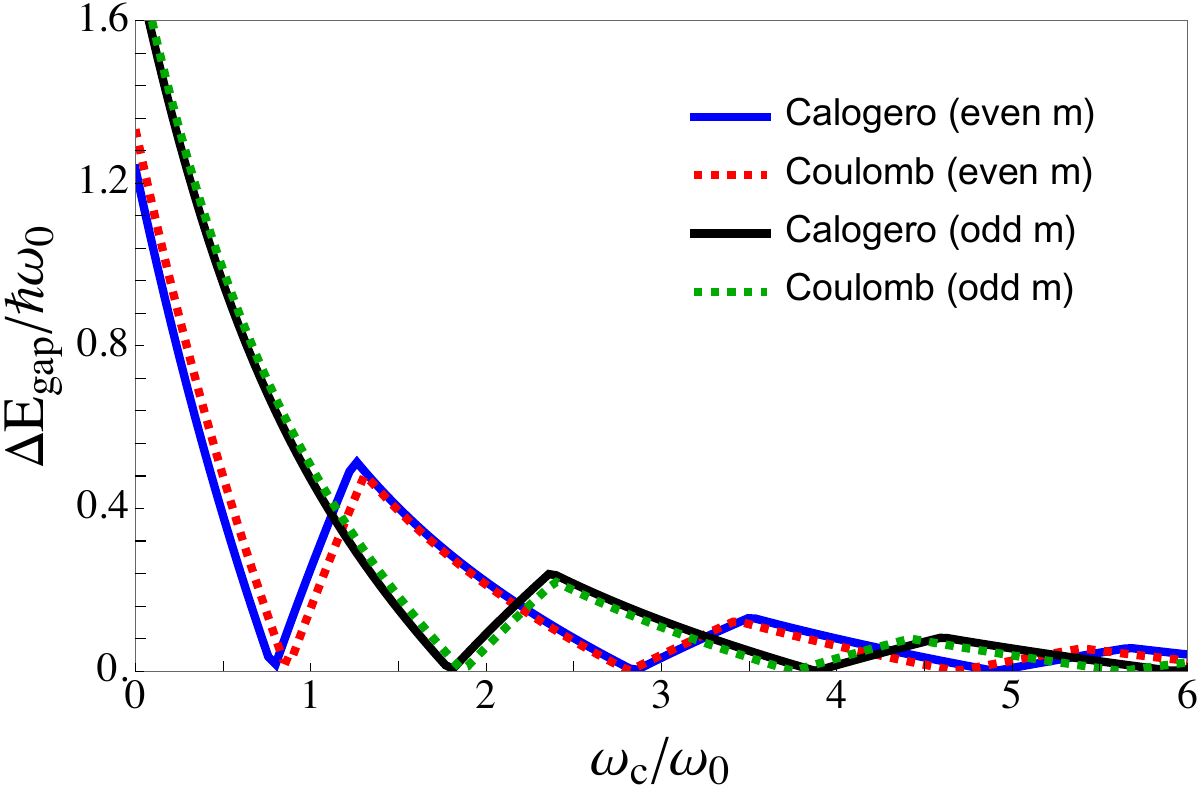}
\vspace{-0.3em}
\caption{Low-energy gap versus magnetic field: Calogero--Coulomb comparison in even/odd $m$ sectors. The plotted quantity is the reduced low-energy gap $\Delta E_{\mathrm{gap}}/(\hbar\omega_0)$, defined as $\Delta E_{\mathrm{gap}}=E_2-E_1$, where $E_1$ and $E_2$ are the two lowest eigenenergies obtained from diagonalizing the relative-motion Hamiltonian at each value of $\omega_c/\omega_0$. Solid curves correspond to the inverse-square (Calogero) interaction ($\eta=2$), while dashed curves correspond to the Coulomb case ($\eta=1$). Results are shown separately for the even-$m$ and odd-$m$ families.}
\label{fig:gapbenchmark}
\vspace{-0.8em}
\end{figure}

Further support for this point is provided by benchmarking the Calogero scaffold against the Coulomb interaction, by comparing the low-energy spectral gap $\Delta E_{\mathrm{gap}}=E_2-E_1$ as a function of $\omega_c/\omega_0$ (Fig.~\ref{fig:gapbenchmark}). The close agreement between the Calogero and Coulomb curves indicates that the qualitative low energy spectral structure, including the locations of near degeneracies and the field dependence of the lowest gap, is robust upon replacing the solvable $1/r^2$ interaction by the more realistic $1/r$ form, within the parameter range relevant to the proposed twisted-light addressing protocol.

\begin{figure}[t]
\centering
\includegraphics[width=\columnwidth]{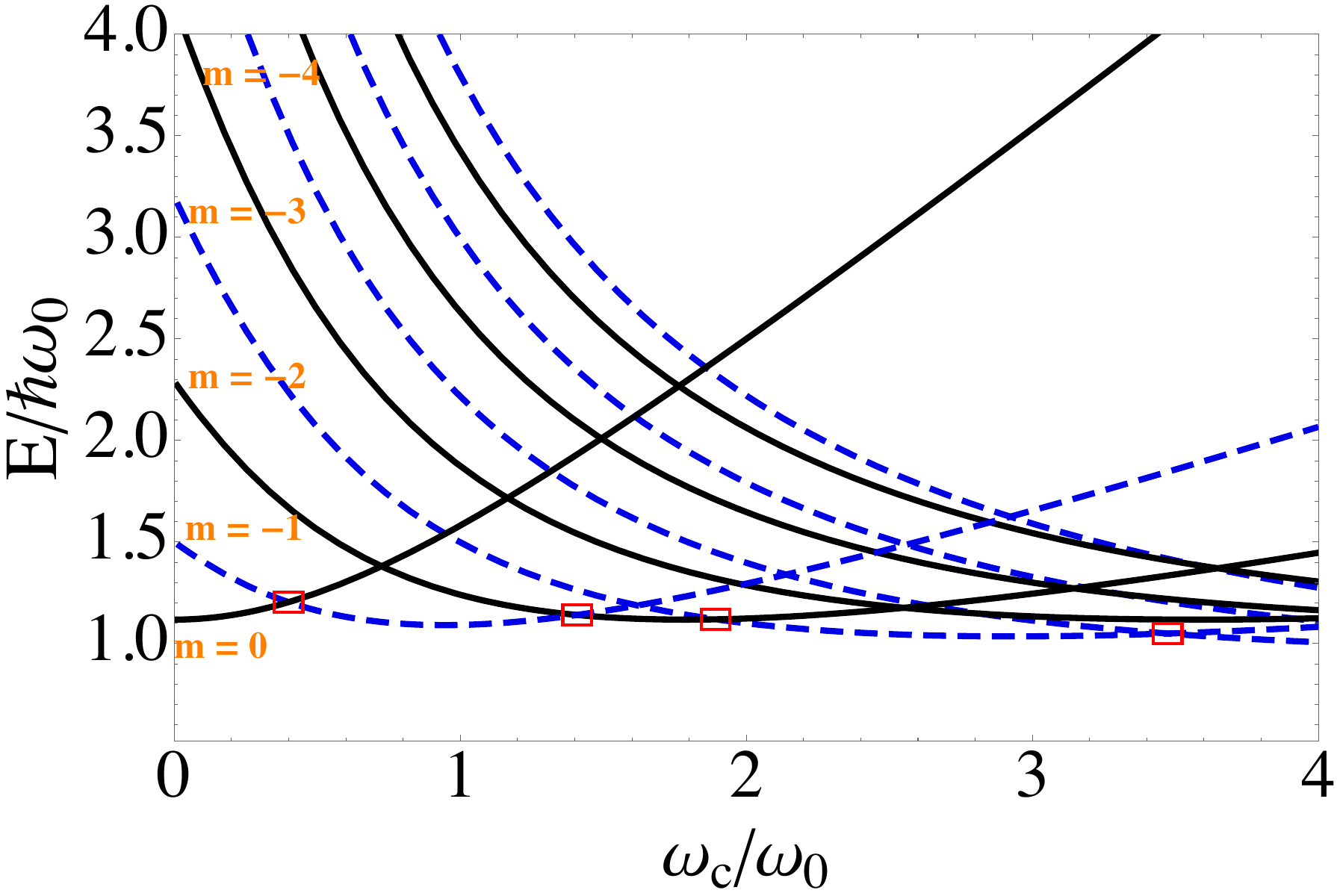}
\vspace{-0.3em}
\caption{Representative relative-motion spectrum and correlation-sector switches. Low-lying relative energies (in units of $\hbar\omega_0$) as a function of the reduced magnetic field $\omega_c/\omega_0$ at fixed interaction strength $\alpha=1.25$. Curves are labelled by the angular-momentum (correlation) sector $m$. Red squares mark the field values at which the ground state switches between neighbouring correlation sectors (level crossings), consistent with the sector boundaries in Fig.~\ref{fig:mGSheatmap}.}
\label{fig:spectrumcrossings}
\vspace{-0.8em}
\end{figure}

To connect the control map in Fig.~\ref{fig:mGSheatmap} to the underlying low-energy structure, Fig.~\ref{fig:spectrumcrossings} shows a representative set of low-lying relative-motion energies as a function of the reduced magnetic field $\omega_c/\omega_0$ at fixed interaction strength $\alpha$. Each branch is labeled by the angular-momentum (correlation) sector $m$, and the red-highlighted points indicate the field values at which the ground state changes from one sector to the next. In other words, the sector boundaries in Fig.~\ref{fig:mGSheatmap} correspond directly to crossings of the lowest-energy branches in Fig.~\ref{fig:spectrumcrossings}, where $E_{\mathrm{rel}}(0,m)$ becomes degenerate with $E_{\mathrm{rel}}(0,m')$. This spectral view makes transparent how tuning $\omega_c/\omega_0$ reorganizes the ordering of the correlation sectors and, in particular, how increasing field progressively favors larger $|m|$. Importantly, these same $m$-labeled ladders are the internal relative states targeted by twisted-light driving through the selection rule $\Delta|m|=\pm(l+\sigma)$ discussed above; Fig.~\ref{fig:spectrumcrossings} therefore provides a concrete spectral reference for identifying operating points and for interpreting the corresponding excitation response shown in Fig.~\ref{fig:excitationspectrumcrossings}.
\subsection{Exact twisted-light matrix elements}

For transitions between $n=0$ Calogero states labelled by the nonnegative sector index $|m|$ (so we write $m\ge 0$ below) driven by twisted light carrying total angular momentum $s=l+\sigma$, the reduced matrix element takes the closed form\cite{JRQ_tutorial}
\begin{equation}
\mathcal{M}^{(\alpha)}_{m\to m+s}
=
\frac{
\Gamma\!\left(\frac{\alpha_m+\alpha_{m+s}+s+2}{2}\right)
}{
\sqrt{\Gamma(\alpha_m+1)\,\Gamma(\alpha_{m+s}+1)}
}.
\label{eq:matrix_element}
\end{equation}
This analytic dependence is not a mathematical luxury: it directly sets the Rabi frequency of a driven qubit and therefore the gate time and fidelity (Section~V).

\section{Quantum computing mapping: correlation-sector qubit and gates}

\subsection{Qubit encoding in correlation sectors}

A computational basis is defined from two odd-$|m|$ relative-motion sectors.
A concrete choice is
\begin{equation}
\begin{split}
\ket{0} &\equiv \ket{m=-3} \quad (\text{sector } |m|=3), \\
\ket{1} &\equiv \ket{m=-5} \quad (\text{sector } |m|=5).
\end{split}
\label{eq:qubit_def}
\end{equation}
which are adjacent under the $s=2$ selection rule.
Throughout, we make the reasonable assumption that the spin-polarized triplet manifold is selected by magnetic field and Zeeman splitting, so the qubit is encoded purely in the orbital correlation sector.
Physically, $\ket{1}$ has a larger correlation hole than $\ket{0}$ and a larger typical relative separation.

This encoding is not (yet) topological: at $N=2$ there is no anyonic braiding, no edge conformal field theory, and no topological degeneracy.
The value is instead that these are \emph{interaction-defined} wavefunctions with distinct correlations, and twisted light addresses them selectively.

\subsection{Single-qubit X rotations: twisted-light Rabi oscillations}

Near resonance with the $\ket{0}\leftrightarrow\ket{1}$ transition and within the rotating-wave approximation, the driven two-level subspace has an effective Hamiltonian of the form
\begin{equation}
H_{\mathrm{eff}}
=
\frac{\hbar\delta}{2}\sigma_z
+
\frac{\hbar\Omega_R}{2}\sigma_x,
\label{eq:Heff}
\end{equation}
where $\delta$ is the detuning and $\Omega_R$ is the Rabi frequency.
Up to known geometric prefactors from the optical field, $\Omega_R$ is proportional to the Calogero matrix element in Eq.~\eqref{eq:matrix_element}:
\begin{equation}
\Omega_R \propto {\cal E}_0\,\left|\mathcal{M}^{(\alpha)}_{3\to 5}\right|.
\label{eq:Rabi_scaling}
\end{equation}
A resonant pulse of duration $t$ implements
\begin{equation}
R_x(\theta)=\exp\!\left(-i\frac{\theta}{2}\sigma_x\right),
\qquad \theta=\Omega_R t,
\label{eq:Rx_gate}
\end{equation}
i.e.\ an $X$ rotation with angle set by pulse area.

\subsection{Single-qubit Z rotations: phase accumulation}

Free evolution for a time $\tau$ between pulses produces a $Z$ rotation through the energy splitting between the two chosen sectors:
\begin{equation}
\hbar\omega_{\mathrm{qubit}}=E_{0,-5}^{\mathrm{rel}}-E_{0,-3}^{\mathrm{rel}}
=\hbar\omega\,(\alpha_{5}-\alpha_{3})-\hbar\omegac,
\label{eq:qubit_freq}
\end{equation}
i.e.
\begin{equation}
\omega_{\mathrm{qubit}}=\omega\left(\sqrt{25+\alpha}-\sqrt{9+\alpha}\right)-\omegac.
\end{equation}
The corresponding phase gate is
\begin{equation}
R_z(\phi)=\exp\!\left(-i\frac{\phi}{2}\sigma_z\right),
\qquad \phi=\omega_{\mathrm{qubit}}\tau.
\label{eq:Rz_gate}
\end{equation}
Because Eq.~\eqref{eq:calogero_energy} is analytic in $\alpha$, the qubit frequency is known in closed form.

\subsection{Universality: compiling arbitrary single-qubit unitaries from Rx and Rz}

In circuit-model language, a \emph{universal} single-qubit gate set is one from which any unitary $U\in SU(2)$ can be synthesized to arbitrary accuracy.
The pair $\{R_x(\theta),R_z(\phi)\}$ is universal: any single-qubit unitary admits an Euler-angle decomposition
\begin{equation}
U = R_z(\phi_3)\,R_x(\theta)\,R_z(\phi_1),
\label{eq:Euler_decomp}
\end{equation}
so in practice a pulse--wait--pulse sequence suffices to implement an arbitrary rotation.
The significance of the $N=2$ TL--Calogero system is that both the pulse-driven angle $\theta=\Omega_R t$ and the free-evolution phase $\phi=\omega_{\mathrm{qubit}}\tau$ are controlled by \emph{analytic} functions of the interaction parameter $\alpha$ (Eqs.~\eqref{eq:Rabi_scaling} and \eqref{eq:calogero_energy}).

\subsection{Readout: twisted-light spectroscopy as a correlation diagnostic}

A key conceptual point is that twisted light provides a natural \emph{state-discrimination channel}:
the same selection rules that enable writing excitations also yield correlation-sensitive absorption lines.\cite{Rodriguez2025}
Spectroscopy is a route toward readout by resolving transitions that originate predominantly from $\ket{0}$ or from $\ket{1}$ (e.g.\ by driving to a higher $|m|$ sector with distinct resonance).
A true projective, single-shot measurement still requires a device-specific detector architecture and signal-to-noise analysis.
This gives the platform a useful write and diagnose symmetry: prepare with one pulse family, interrogate with the same optical channel.

\begin{figure}[tbh]
\centering
\includegraphics[width=\columnwidth]{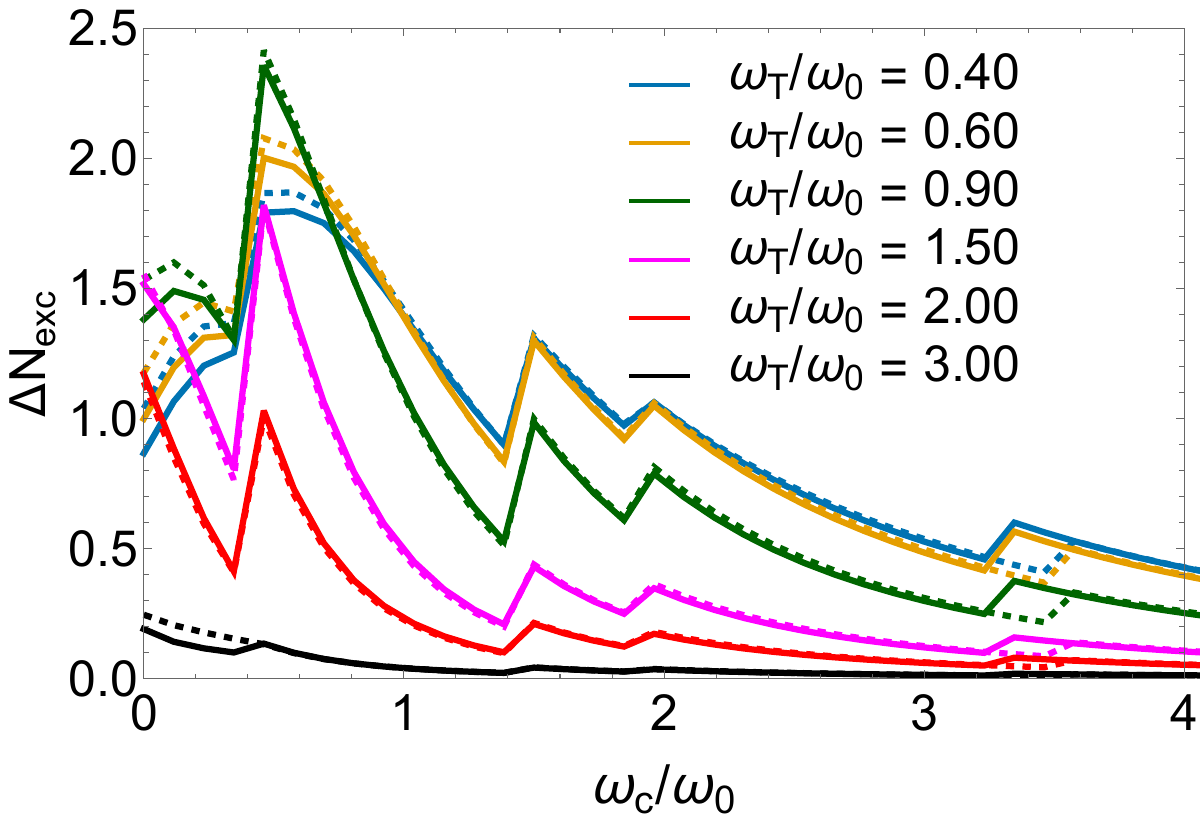}
\vspace{-0.3em}
\caption{Twisted-light excitation response as a correlation-sensitive readout proxy. Mean net change in the number of chiral excitations, $\langle\Delta N_{\mathrm{exc}}\rangle$, plotted versus $\omega_c/\omega_0$ for several fixed pulse central frequencies $\omega_T/\omega_0$ (as indicated). Solid (dashed) curves correspond to the Coulomb (Calogero) interaction model. Pronounced variations of $\langle\Delta N_{\mathrm{exc}}\rangle$ track resonance conditions and symmetry-allowed transitions under the selection rule $\Delta|m|=\pm(l+\sigma)$, providing an experimentally oriented diagnostic of the underlying correlation-sector structure.}
\label{fig:excitationspectrumcrossings}
\vspace{-0.8em}
\end{figure}

Beyond providing a state-selective writing channel, twisted light also offers a natural route to readout via spectroscopy. To make this explicit, in Fig.~\ref{fig:excitationspectrumcrossings} we plot the excitation response in terms of the mean net change in the number of chiral excitations, $\langle \Delta N_{\mathrm{exc}} \rangle$, evaluated from the pulse-driven dynamics and taken in practice as a proxy for absorption at the chosen central frequency $\omega_T$.

Sweeping the magnetic field at fixed $\omega_T/\omega_0$ produces pronounced step-like features and sharp variations in $\langle \Delta N_{\mathrm{exc}} \rangle$, which can be traced directly to the underlying level structure in Fig.~\ref{fig:spectrumcrossings}: when the ground state changes correlation sector (or becomes nearly degenerate with the first accessible excitation in the same parity family), resonant channels open or close under the twisted-light selection rule $\Delta|m|=\pm(l+\sigma)$, leading to enhanced or suppressed excitation. In this way, Figs.~\ref{fig:spectrumcrossings} and~\ref{fig:excitationspectrumcrossings} together illustrate the central write-diagnose symmetry of the platform: the same structured-light channel that addresses internal correlation ladders also provides correlation-sensitive spectral fingerprints suitable for distinguishing sectors and calibrating operating points.

\subsection{Why the Calogero parameter matters for gate design}

The interaction parameter $\alpha$ enters both the spectrum (Eq.~\eqref{eq:calogero_energy}) and the matrix element (Eq.~\eqref{eq:matrix_element}).
Hence, it acts as a continuous ``knob'' that tunes:
\begin{itemize}
\item the qubit frequency $\omega_{\mathrm{qubit}}$ (phase gate rate),
\item the Rabi frequency $\Omega_R$ (pulse time),
\item leakage and selectivity, by controlling level spacings to nearby sectors.
\end{itemize}
In realistic dots, $\alpha$ is not literally an inverse-square coupling, but it can be viewed as an effective parameter capturing screening and confinement; the analytically solvable point provides guidance for robust operating regimes.\cite{Rodriguez2025}

\subsection{Leakage and anharmonicity: why an interacting ladder can help gate fidelity}

A practical qubit must behave approximately as a two-level system under the applied drive.
Here the relevant leakage channels are transitions from $\ket{0}\equiv |m|=3$ and $\ket{1}\equiv |m|=5$ into neighboring sectors such as $|m|=1$ or $|m|=7$.
A helpful feature of the Calogero spectrum is that it is \emph{anharmonic} in $m$ because $\alpha_m=\sqrt{m^2+\alpha}$.
For a TL mode carrying $s$ units of angular momentum, the transition frequency between $n=0$ sectors is
\begin{equation}
\begin{split}
\omega_{|m|\to |m|+s} &= \frac{E^{\mathrm{rel}}_{0,-(|m|+s)}-E^{\mathrm{rel}}_{0,-|m|}}{\hbar} \\
&= \omega\left(\alpha_{|m|+s}-\alpha_{|m|}\right)-\frac{s\omegac}{2}.
\end{split}
\label{eq:transition_freq}
\end{equation}
For the qubit choice in Eq.~\eqref{eq:qubit_def} with $s=2$, the \emph{anharmonicity} that controls off-resonant leakage to $|m|=7$ is
\begin{equation}
\mathcal{A}\equiv \omega_{5\to 7}-\omega_{3\to 5}
=\omega\left[(\alpha_7-\alpha_5)-(\alpha_5-\alpha_3)\right].
\label{eq:anharmonicity}
\end{equation}
Because $\mathcal{A}$ is an analytic function of $\alpha$, the solvable limit provides a direct way to identify parameter regimes where $\mathcal{A}$ is large compared to the Rabi rate $\Omega_R$, suppressing leakage.
This is the same ``anharmonicity engineering'' logic used in transmons, but here it arises from \emph{correlations} rather than Josephson physics.

\section{Toward a photonic control bus and multi-qubit operations}

\subsection{Write, read, scale: a minimal control narrative}

Table~\ref{tab:WRS} summarizes the operational viewpoint suggested by the physics above.

\begin{table*}[t]
\centering
\caption{Twisted light as a control layer for correlation-sector qubits.}
\label{tab:WRS}
\renewcommand{\arraystretch}{1.15}
\begin{tabular}{@{}p{0.10\textwidth}p{0.40\textwidth}p{0.44\textwidth}@{}}
\toprule
\textbf{Role} & \textbf{Physical action} & \textbf{Quantum-information meaning} \\
\midrule
WRITE & Resonant TL pulse (choose $l,\sigma$, pulse area) & Implements $R_x(\theta)$ between correlation sectors. \\
READ & TL spectroscopy of correlation-sensitive lines & Distinguishes correlation sectors; projects onto $\ket{0}$ or $\ket{1}$. \\
SCALE & SLM steering + vorticity selection & Optical addressing of many dots without dense wiring; mode selectivity sets $\Delta|m|$. \\
\bottomrule
\end{tabular}
\end{table*}

\subsection{Addressability: diffraction limit and near-field requirements at THz frequencies}
\label{sec:addressability}

The ``SCALE'' line in Table~\ref{tab:WRS} intentionally highlights an attractive idea: a spatial light modulator (SLM) can steer a beam to select a dot, while the OAM/polarization choice selects $\Delta|m|$ and therefore the internal transition.
However, for the intraband orbital transitions emphasized here, the natural frequencies are typically in the THz range (Sec.~IX), corresponding to free-space wavelengths $\lambda\sim 0.3$--$3~\mathrm{mm}$.
A diffraction-limited far-field spot size is therefore orders of magnitude larger than typical lithographic dot spacings (tens to hundreds of nm, or $\sim\mu$m-scale pitches in larger arrays).
This mismatch is the central scalability challenge for the ``photonic control bus'' vision and should be stated explicitly.

There are several physically plausible routes forward, all of which fit naturally into the META (metamaterials/photonics) scope:
(i) \emph{Proof-of-principle sparse arrays}: begin with isolated dots separated by tens--hundreds of $\mu$m so that a THz focus can address them individually.
(ii) \emph{On-chip near-field localization}: integrate each dot with a local THz antenna, resonator, or metasurface element that converts an incident beam into a tightly confined near field at the dot location (subwavelength confinement is routine in resonant near-field structures even when $\lambda$ is macroscopic).
(iii) \emph{Waveguide delivery}: route THz/mid-IR power on chip to selected sites, using the optical beam mainly as a global clock or for frequency selectivity.
(iv) \emph{Frequency-upshifted designs}: use stronger confinement and/or different materials to push the relevant orbital splittings into the mid-IR, where $\lambda$ is $\sim 10~\mu$m and far-field addressing becomes less extreme.

In short: mode selectivity via OAM is a genuine asset, but \emph{spatial} selectivity at THz wavelengths will likely require near-field or structured-photonic engineering rather than simple far-field focusing.

\subsection{A concrete two-qubit entangling gate mechanism}

A scalable architecture requires at least one entangling two-qubit gate.
For two nearby dots $A$ and $B$, the inter-dot Coulomb interaction depends on the charge distribution in each dot.
Because different correlation sectors have different spatial extent, the interaction energy is state dependent.
A clean analytic diagnostic of this size difference is the exact Calogero expectation value\cite{JRQ_tutorial}
\begin{equation}
\langle r^2\rangle_{m,\alpha}=4\ellzero^2(\alpha_m+1),
\qquad \alpha_m=\sqrt{m^2+\alpha}.
\label{eq:r2_expectation}
\end{equation}
This motivates an effective $ZZ$ coupling model:
\begin{equation}
H_{ZZ}=\frac{\hbar J}{4}\,\sigma_z^{(A)}\sigma_z^{(B)},
\qquad
J\equiv\frac{U_{11}+U_{00}-U_{10}-U_{01}}{\hbar},
\label{eq:ZZ_coupling}
\end{equation}
where $U_{ij}$ is the electrostatic energy when dot $A$ is in $\ket{i}$ and dot $B$ in $\ket{j}$.

\subsubsection{Closed-form estimate of $J$ from a long-distance multipole expansion}
\label{sec:J_multipole}

Equation~\eqref{eq:ZZ_coupling} defines the entangling rate $J$, but it is useful to have an explicit estimate.
Assume two \emph{non-overlapping} dots whose centers are separated by an in-plane distance $D$ (typically $D\sim 200$--$500~\mathrm{nm}$ in gated devices\cite{Kouwenhoven2001}).
Let the electron positions in dot $A$ be $\mathbf{r}_{A,1}=\mathbf{R}_A+\mathbf{r}_A/2$ and $\mathbf{r}_{A,2}=\mathbf{R}_A-\mathbf{r}_A/2$, where $\mathbf{r}_A$ is the internal (relative) coordinate; similarly for dot $B$.
The inter-dot interaction is then
\begin{equation}
V_{AB}=\frac{e^2}{4\pi\varepsilon}\sum_{a=1}^2\sum_{b=1}^2
\frac{1}{\big|\mathbf{D}+\mathbf{r}_{A,a}-\mathbf{r}_{B,b}\big|},
\end{equation}
where $\mathbf{D}\equiv \mathbf{R}_B-\mathbf{R}_A$.
For $D\gg \sqrt{\langle r^2\rangle}$ one may expand $1/|\mathbf{D}+\boldsymbol\delta|$ in powers of $\boldsymbol\delta/D$ and then average over the rotationally symmetric internal states (for which $\langle \mathbf{r}\rangle=\mathbf{0}$).
A derivation can be obtained by Taylor expanding $1/|\mathbf{D}+\boldsymbol\delta|$ and using isotropic angular averages; equivalently, it is the leading quadrupole--quadrupole contribution for planar charge distributions. The result is: 
\begin{equation}
\begin{split}
U_{ij} \approx \frac{e^2}{4\pi\varepsilon} \biggl[ &\frac{4}{D} + \frac{\langle r^2\rangle_i+\langle r^2\rangle_j}{4D^3} \\
&+ \frac{9}{64}\frac{\langle r^2\rangle_i\,\langle r^2\rangle_j}{D^5} + \cdots \biggr],
\end{split}
\label{eq:Uij_multipole}
\end{equation}
where $\langle r^2\rangle_i$ denotes the internal size in dot $A$ when it is in $\ket{i}$, and $\langle r^2\rangle_j$ is the corresponding size in dot $B$.
The $D^{-3}$ term produces only \emph{single-qubit} frequency shifts (it is additive in $i$ and $j$ and cancels in $U_{11}+U_{00}-U_{10}-U_{01}$).
The leading \emph{entangling} contribution is the $D^{-5}$ cross-term, giving
\begin{equation}
J \;\approx\; \frac{1}{\hbar}\frac{e^2}{4\pi\varepsilon}\frac{9}{64}\,
\frac{\big(\Delta\langle r^2\rangle\big)^2}{D^5},
\qquad 
\Delta\langle r^2\rangle \equiv \langle r^2\rangle_{1}-\langle r^2\rangle_{0}.
\label{eq:J_estimate}
\end{equation}
Using the exact Calogero size formula in Eq.~\eqref{eq:r2_expectation},
$\Delta\langle r^2\rangle = 4\ellzero^2(\alpha_5-\alpha_3)$, this becomes
\begin{equation}
J \;\approx\; \frac{1}{\hbar}\frac{e^2}{4\pi\varepsilon}\frac{9}{4}\,
\frac{\ellzero^4(\alpha_5-\alpha_3)^2}{D^5}.
\label{eq:J_estimate_l0}
\end{equation}
For representative GaAs parameters ($\varepsilon=\kappa\varepsilon_0$ with $\kappa\simeq 12.9$ and $\ellzero\simeq 10~\mathrm{nm}$; Sec.~IX) and $\alpha_5-\alpha_3\simeq 2$ (noninteracting limit), Eq.~\eqref{eq:J_estimate} gives $J/(2\pi)\sim 0.08$--$8~\mathrm{MHz}$ for $D=500$--$200~\mathrm{nm}$, corresponding to a controlled-phase time $t_{\mathrm{CZ}}=\pi/J\sim 6~\mu\mathrm{s}$--$70~\mathrm{ns}$.
The steep $D^{-5}$ scaling makes this gate \emph{highly tunable}: bringing dots closer or using a larger effective size difference (larger $\Delta\langle r^2\rangle$) rapidly enhances $J$.

Free evolution for time $t=\pi/J$ implements a controlled-phase (CZ-equivalent) gate up to single-qubit $Z$ rotations.

This mechanism is not topological---it is analogous to capacitive $ZZ$ coupling in other platforms---but it is concrete and testable, and its distinctive feature is that the basis states are \emph{correlation sectors} written and read optically by twisted light.

\subsection{Coupled $N=2$ dots as a two-qubit module: analytic interdot entanglement and an $XY$ gate channel}
\label{sec:coupledN2dots}

\paragraph{Connecting single-dot and coupled-dot ``magic-number'' transitions.}
In an isolated, spin-polarized $N=2$ parabolic dot, the low-energy branch is organized by odd relative angular momenta $|m|=1,3,5,\dots$; as the magnetic field is increased the ground state typically steps through this odd-$|m|$ ladder (the $N=2$ analogue of the well-known $N=3$ magic-number sequence $L=3k$).
For two dots, the uncoupled spectrum is the direct product $\ket{m_A,m_B}$ of the \emph{same} single-dot ladders.
Coulomb-range interdot interactions then reshape these inherited ``magic'' sectors by creating resonant anticrossings and entangled eigenstates when two product states become nearly degenerate.
The Benjamin--Johnson selection rule makes this concrete: the dominant mixing occurs between
$\ket{m,m+2}$ and $\ket{m+2,m}$, i.e.\ the same $\Delta|m|=2$ step that twisted light addresses within a single dot.
In this sense, coupled-dot ``magic-number transitions'' are not a separate phenomenon: they are the single-dot magic ladder \emph{promoted} to a two-site module, where interdot interactions convert near-degeneracies into exchange-like ($XY$) entangling dynamics.

The capacitive $ZZ$ mechanism in Sec.~\ref{sec:J_multipole} is the simplest entangling interaction to model, but it is not the only one available when two dots are brought into Coulomb range.
In fact, a qualitatively different entangling channel---a resonant, exchange-like \emph{$XY$ coupling} between $\ket{01}$ and $\ket{10}$---appears already in early analytic work on coupled few-electron dots with $N=2$ electrons \emph{per dot}.
This is directly relevant to the present proposal because our qubit basis states $\ket{0}\equiv\ket{|m|=3}$ and $\ket{1}\equiv\ket{|m|=5}$ differ by $\Delta|m|=2$, exactly the ``quantum'' transferred between dots in those analytic coupled-dot solutions.

\subsubsection{Benjamin--Johnson multiple-dot model: entanglement without tunneling}

Benjamin and Johnson introduced an analytically solvable model of $P$ colinear, two-dimensional quantum dots, each containing two electrons, arranged vertically with interdot separation $s$.\cite{BenjaminJohnson1995PRB}
Interdot tunneling is neglected; adjacent dots are coupled purely by the two-body electron--electron interaction, an ``optical wiring'' channel that can exist even in the absence of any single-particle tunneling term.\cite{BenjaminJohnson1995PRB}
They take an inverse-square interaction (Calogero form) both within each dot and between adjacent dots,
\begin{align}
V_{\mathrm{intra}} &= \sum_{a=1}^{P}\frac{\beta}{\big|\mathbf{r}_{a,1}-\mathbf{r}_{a,2}\big|^2},\\
V_{\mathrm{inter}} &= \sum_{a=2}^{P}\sum_{i,j=1}^{2}\frac{\beta}{\big|\mathbf{r}_{a,i}-\mathbf{r}_{a-1,j}\big|^2+s^2},
\end{align}
and then Taylor-expand $V_{\mathrm{inter}}$ under the physically transparent assumption that typical in-plane separations are smaller than $s$.\cite{BenjaminJohnson1995PRB}
To order $P r^2/s^4$ the coupled-dot problem remains exactly solvable; higher-order terms (schematically $P r^4/s^6$) are then treated by analytic perturbation theory.\cite{BenjaminJohnson1995PRB}

For $P=2$ identical dots, the low-lying eigenstates can be labeled as direct products $\ket{m_1,m_2}$, where $m_a$ is the relative angular momentum quantum number of the two-electron state in dot $a$.\cite{BenjaminJohnson1995PRB}
At the exactly solvable order ($\propto r^2/s^4$), the pair $\ket{m,m+2}$ is degenerate with $\ket{m+2,m}$.
The next-order interdot perturbation then mixes these states \emph{selectively}:
in the nearly-degenerate subspace, the perturbation only connects $\ket{m_a,m_b}$ to $\ket{m_c,m_d}$ if $m_a-m_c=m_d-m_b=0$ or $\pm 2$.\cite{BenjaminJohnson1995PRB}
As a result, the relevant eigenstates are entangled superpositions of the form
\begin{equation}
\ket{\Psi_{\pm}} = \frac{1}{\sqrt{2}}\Big(\ket{m+2,m}\pm \ket{m,m+2}\Big) \label{eq:entangled_mplus2}
\end{equation}
(identical dots, on resonance) and for unequal dot sizes (weakly nondegenerate case) they become $(p\ket{m+2,m}+q\ket{m,m+2})$ with $p\neq q$ set by the detuning.\cite{BenjaminJohnson1995PRB}
Benjamin and Johnson show that these entangled states can become \emph{ground states} over finite $B$-field windows, and interpret their formation as a resonance phenomenon closely analogous to Förster energy transfer; in an array, the corresponding ``Frenkel exciton'' can propagate and form an exciton band.\cite{BenjaminJohnson1995PRB}

\subsubsection{Mapping to the correlation-sector qubit: an $\mathrm{iSWAP}/\sqrt{\mathrm{iSWAP}}$ entangler}

This structure maps almost one-to-one onto the present qubit encoding.
Take two dots $A,B$ with local basis $\ket{0}\equiv\ket{|m|=3}$ and $\ket{1}\equiv\ket{|m|=5}$.
Then the two-qubit states
\begin{equation}
\ket{01}\equiv\ket{3,5},\qquad \ket{10}\equiv\ket{5,3}
\end{equation}
are precisely the $(m,m+2)$ and $(m+2,m)$ pair with $m=3$ in Eq.~\eqref{eq:entangled_mplus2}.
The coupled-dot perturbation therefore generates an effective \emph{exchange} Hamiltonian in the $\{\ket{01},\ket{10}\}$ subspace:
\begin{equation}
\begin{split}
H_{XY} &= \hbar J_{\mathrm{ex}}\Big(\ket{01}\!\bra{10}+\ket{10}\!\bra{01}\Big) \\
&= \frac{\hbar J_{\mathrm{ex}}}{2}\Big(\sigma_x^{(A)}\sigma_x^{(B)}+\sigma_y^{(A)}\sigma_y^{(B)}\Big),
\end{split}
\label{eq:H_XY}
\end{equation}
with $J_{\mathrm{ex}}$ set by the relevant interdot matrix element (the same perturbative order that produces the entangled eigenstates).
On exact resonance (identical dots), the eigenstates are the Bell states $\ket{\Psi_{\pm}}=(\ket{01}\pm\ket{10})/\sqrt{2}$ with splitting $2\hbar J_{\mathrm{ex}}$.
Free evolution under $H_{XY}$ implements an entangling gate:
\begin{align}
U(t)\ket{01}&=\cos(J_{\mathrm{ex}}t)\ket{01}-i\sin(J_{\mathrm{ex}}t)\ket{10},\\
U(t)\ket{10}&=\cos(J_{\mathrm{ex}}t)\ket{10}-i\sin(J_{\mathrm{ex}}t)\ket{01}.
\end{align}
Thus $t=\pi/(2J_{\mathrm{ex}})$ yields an $\text{iSWAP}$ (up to single-qubit phases), and $t=\pi/(4J_{\mathrm{ex}})$ yields a $\sqrt{\mathrm{iSWAP}}$ which is maximally entangling.
Crucially, the same coupled-dot analysis provides a \emph{control knob}: detuning the dots (e.g.\ via slightly different confinement strengths) moves the $\ket{01}$ and $\ket{10}$ energies apart, suppressing the mixing (the ``quadratic Stark'' analog in Ref.~\cite{BenjaminJohnson1995PRB}).
In quantum-information language, this is an on/off switch for the $XY$ entangler, complementary to the always-on but tunable $ZZ$ coupling of Sec.~\ref{sec:J_multipole}.

\subsubsection{Johnson--Benjamin double-layer model: analytic four-electron precursor to bilayer FQHE}

A closely related analytic benchmark is provided by Johnson and Benjamin's exactly solvable \emph{double-layer} (vertical) model with two electrons per layer.\cite{JohnsonBenjamin1995JPCM}
They consider electrons confined in each layer by a 2D parabolic potential in a perpendicular magnetic field, neglect inter-layer tunneling, and include both intra- and inter-layer inverse-square interactions.
By introducing center-of-mass and relative coordinates within each layer and between the layer centers of mass, and then Taylor-expanding the inter-layer interaction under the assumption that the typical in-plane separation is less than the layer separation $s$, the Hamiltonian becomes exactly solvable once the expansion is truncated at quadratic order.\cite{JohnsonBenjamin1995JPCM}
In particular, truncation at the second term yields an additive constant plus a sum of four decoupled solvable Hamiltonians,
\begin{equation}
\begin{split}
H &= \frac{4\beta}{s^2} + \sum_{i=1}^{4} H_i(u_i,q_i), \\
\omega_1 &= \omega_0, \quad \omega_2 = \omega_3 = \Big(\omega_0^2 - \frac{4\beta}{m^\ast s^4}\Big)^{1/2}, \\
\omega_4 &= \Big(\omega_0^2 - \frac{8\beta}{m^\ast s^4}\Big)^{1/2}.
\end{split}
\label{eq:JB_H_decomposition}
\end{equation}
where $H_1$ and $H_4$ are single-particle Fock--Darwin problems and $H_2,H_3$ are two-body (Calogero) problems for the intra-layer relative coordinates.\cite{JohnsonBenjamin1995JPCM}
The resulting ground-state sequence in magnetic field is $\ket{m_2,m_3}=\ket{1,1},\ket{3,3},\ket{5,5},\dots$ with odd $m_{2,3}$ set by antisymmetry within each layer.\cite{JohnsonBenjamin1995JPCM}
Johnson and Benjamin further define an effective small-$N$ filling factor per layer $\nu_a\simeq 1/m_a$, giving a total sequence $\nu=\nu_1+\nu_2=2,\,2/3,\,2/5,\dots$ that is consistent with the observed bilayer fractions for sufficiently separated layers.\cite{JohnsonBenjamin1995JPCM}
They also emphasize that the analytically obtained correlated wavefunctions differ fundamentally from the standard Jastrow/Greek--Roman $(mmn)$ forms used in bilayer FQHE modeling.\cite{JohnsonBenjamin1995JPCM}

For the present proposal, Eq.~\eqref{eq:JB_H_decomposition} is useful in two ways.
First, it provides a closed-form dependence of mode frequencies on the geometric separation $s$, making it an analytic ``calibration model'' for how vertical coupling renormalizes energy scales (and hence how interdot coupling shifts qubit splittings and detunings).
Second, it highlights a genuinely multi-dot feature absent in the single-dot problem: an additional collective coordinate (the relative coordinate between layer centers of mass) with its own tunable frequency $\omega_4(s)$.
In an engineered device, this degree of freedom can be viewed as a candidate mediator mode for coupling qubits beyond pure static $ZZ$ shifts.

\subsubsection{How this strengthens the twisted-light quantum-computing proposal}

Taken together, these two analytic coupled-dot benchmarks strengthen and extend the QC narrative in three concrete ways:

\begin{enumerate}
\item \textbf{They provide an explicit, symmetry-selective $XY$ entangler for the same $\Delta|m|=2$ qubit ladder.}
The $\ket{3,5}\leftrightarrow\ket{5,3}$ mixing predicted for coupled $N=2$ dots is precisely the exchange channel needed for $\text{iSWAP}$-type two-qubit gates, complementing the capacitive $ZZ$ gate in Sec.~\ref{sec:J_multipole}.

\item \textbf{They suggest an ``exciton transport'' architecture for dot arrays.}
In the Benjamin--Johnson picture, a localized change of $m$ on one dot can resonantly propagate as a Frenkel exciton band in a chain.\cite{BenjaminJohnson1995PRB}
In our language, this is propagation of a local correlation-sector excitation.
Twisted light provides the natural way to \emph{write} such an excitation on a selected dot and \emph{read} it out spectroscopically at another location.

\item \textbf{They supply analytic $s$-dependence and mode structure for vertically coupled modules.}
The Johnson--Benjamin decomposition makes clear which collective coordinates are activated by vertical coupling and how their frequencies scale with $s$.\cite{JohnsonBenjamin1995JPCM}
This is exactly the kind of analytic handle needed to design detuning-based on/off control of interdot exchange and to anticipate extra spectator modes that may need to be dynamically decoupled.
\end{enumerate}

In short: the coupled-dot literature already contains an analytic demonstration that two $N=2$ dots can generate strong, interaction-mediated \emph{entangled} eigenstates without tunneling.
Twisted light adds the missing ingredient for quantum information: a structured, sector-selective optical handle to coherently address the internal ladders and thereby turn those coupled-dot correlations into a programmable gate set.

\subsection{Comparison with established qubit platforms}

The correlation-sector qubit shares features with several established platforms while introducing genuinely new elements: (i)~an interacting many-body basis, and (ii)~structured-light control of internal motion.
Table~\ref{tab:comparison} provides a compact comparison.

\begin{table*}[t]
\centering
\caption{Context: correlation-sector qubit compared to common platforms (qualitative).}
\label{tab:comparison}
\renewcommand{\arraystretch}{1.15}
\small
\begin{tabular}{p{0.14\textwidth}p{0.26\textwidth}p{0.26\textwidth}p{0.26\textwidth}}
\toprule
\textbf{Feature} & \textbf{$N=2$ correlation-sector qubit} & \textbf{Superconducting transmon} & \textbf{Trapped ion} \\
\midrule
Basis states & Relative-motion sectors $|m|=3,5$ (correlations) & Weakly anharmonic oscillator levels & Hyperfine/optical atomic levels \\
Single-qubit drive & Twisted light pulse ($l,\sigma$) & Microwave pulse & Laser pulse \\
Native $X$ gate & TL Rabi oscillation (Eq.~\eqref{eq:Rx_gate}) & Microwave Rabi & Raman/direct Rabi \\
Native $Z$ gate & Free evolution (Eq.~\eqref{eq:Rz_gate}) & Detuning/flux bias & Free evolution \\
Anharmonicity & Correlation-induced $\alpha_m=\sqrt{m^2+\alpha}$ & Josephson nonlinearity & Atomic structure \\
Analytic gate parameters & \textbf{Exact} in Calogero limit (Eqs.~\eqref{eq:calogero_energy}, \eqref{eq:matrix_element}) & Typically calibrated numerically & Exact (atomic theory) \\
Two-qubit coupling idea & State-dependent Coulomb $ZZ$ (Eq.~\eqref{eq:ZZ_coupling}) & Capacitive/inductive coupling & Shared motional modes \\
Topological upgrade path & $N\ge 3$ FQHE droplets and quasiholes & None intrinsic & None intrinsic \\
\bottomrule
\end{tabular}
\end{table*}


\section{Noise channels and robustness considerations}

Because the encoding in Eq.~\eqref{eq:qubit_def} uses different \emph{charge-correlation} sectors, the dominant decoherence mechanisms are expected to resemble those of ``charge-like'' qubits rather than pure spin qubits.
This does not invalidate the platform; it clarifies what must be engineered and what ``robustness'' realistically means at the $N=2$ stage.

\subsection{Control errors: pulse area, detuning, and mode purity}

Twisted-light gates rely on resonant driving, so standard coherent-control considerations apply:
(i)~pulse area errors ($\theta\neq \Omega_R t$),
(ii)~detuning errors ($\delta\neq 0$ in Eq.~\eqref{eq:Heff}),
and (iii)~imperfect OAM/helicity mode purity (admixture of unwanted $s$ values).
The advantage of the analytic framework is that these errors can be propagated through closed-form expressions for the gate unitary and (in principle) optimized using composite pulses and robust control sequences, without relying purely on black-box calibration.

\subsection{Environmental decoherence: phonons and electrostatic noise}

In a semiconductor dot, coupling to acoustic phonons and to charge noise in nearby gates and impurities will induce relaxation and dephasing between correlation sectors.
Since $\ket{0}$ and $\ket{1}$ differ in spatial extent, they couple differently to fluctuating electrostatic potentials, producing dephasing analogous to that of charge qubits.
A practical strategy is therefore to operate in regimes where:
(i)~the energy splitting $\hbar\omega_{\mathrm{qubit}}$ is large compared to $k_BT$ (suppressing thermal mixing),
and (ii)~the relevant transitions are well separated from leakage channels (large anharmonicity $\mathcal{A}$ in Eq.~\eqref{eq:anharmonicity}).

\subsection{What ``robust'' can mean in this context}

At $N=2$, ``robust'' should not be interpreted as topological fault tolerance.
Instead, it refers to two experimentally useful features already demonstrated in the spectroscopy setting:
(i)~twisted light targets internal excitations that are symmetry forbidden to dipole probes (a clean signal channel), and
(ii)~the qualitative chiral fingerprints persist for a range of interaction forms from screened to Coulomb.\cite{Rodriguez2025}
These properties motivate the $N=2$ dot as a \emph{calibration-grade} gate primitive and as a stepping stone to $N\ge 3$ droplets where true topological protection may be approached.

\subsection{Decoherence: a quantitative dephasing estimate and the key engineering challenge}
\label{sec:decoherence_estimate}

A realistic assessment must confront decoherence.
Our encoding uses \emph{orbital correlation sectors} ($|m|=3$ and $|m|=5$), so it is ``charge-like'' in the sense that the energy depends on the spatial distribution.
However, a definite-$m$ state in an ideal circular dot is \emph{rotationally symmetric} and satisfies $\langle \mathbf{r}\rangle=\mathbf{0}$, so it carries \emph{no permanent in-plane dipole moment}.
As a result, spatially uniform electric-field noise couples primarily to the center-of-mass (Kohn) mode rather than directly dephasing the internal qubit.
The leading internal sensitivity is instead \emph{quadrupolar}: fluctuations of the confining curvature (or equivalently, field gradients) shift the two qubit levels differently because $\langle r^2\rangle$ differs.

A simple analytic estimate follows from a ``curvature-noise'' model in which gate-voltage noise induces a small fluctuation $\delta\omega_0$ of the bare harmonic confinement.
Since $\omega^2=\omega_0^2+\omega_c^2/4$, the relative Hamiltonian contains a term $(\mu\omega^2/2)\,r^2$ with $\partial H/\partial\omega_0=\mu\omega_0 r^2$.
By the Hellmann--Feynman theorem,
\begin{equation}
\frac{\partial E_m}{\partial \omega_0}=\mu\omega_0\,\langle r^2\rangle_{m,\alpha},
\qquad
\Rightarrow\qquad
\frac{\partial \omega_{\mathrm{qubit}}}{\partial \omega_0}
=\frac{\mu\omega_0}{\hbar}\,\Delta\langle r^2\rangle,
\label{eq:domega_qubit_domega0}
\end{equation}
with $\Delta\langle r^2\rangle\equiv \langle r^2\rangle_{|m|=5}-\langle r^2\rangle_{|m|=3}=4\ellzero^2(\alpha_5-\alpha_3)$ from Eq.~\eqref{eq:r2_expectation}.
For quasi-static Gaussian noise with rms amplitude $\sigma_{\omega_0}$, the rms fluctuation of the qubit splitting is
\begin{equation}
\sigma_{\omega_{\mathrm{qubit}}}=
\left|\frac{\partial \omega_{\mathrm{qubit}}}{\partial \omega_0}\right|\sigma_{\omega_0}
=
\frac{\mu\omega_0}{\hbar}\,\Delta\langle r^2\rangle\,\sigma_{\omega_0},
\label{eq:T2star_estimate}
\end{equation}
so that $T_2^*\sim \sigma_{\omega_{\mathrm{qubit}}}^{-1}$.
Equation~\eqref{eq:T2star_estimate} cleanly identifies the levers that improve dephasing: suppress curvature noise ($\sigma_{\omega_0}$), reduce sensitivity via large magnetic field ($\omega_0/\omega\ll 1$ so gate voltages weakly perturb $\omega$), or engineer a smaller $\Delta\langle r^2\rangle$ (at the cost of reduced $J$ in Sec.~\ref{sec:J_multipole}).

To connect with experiment, charge qubits in semiconductor double quantum dots typically exhibit $T_2^*$ in the ns range when operated away from optimized ``sweet spots'' (e.g., a decoherence time $\sim 1~\mathrm{ns}$ in early GaAs charge-qubit oscillations\cite{Hayashi2003}, and a maximum coherence time $\sim 7~\mathrm{ns}$ at the charge-degeneracy point consistent with a $1/f$ noise model\cite{Petersson2010}).
An important mitigating feature of our \emph{single-dot, circular} encoding is that the internal qubit is not a dipole qubit; in an ideal harmonic dot the leading charge-noise coupling can be strongly suppressed\cite{Gamble2012}.
Nevertheless, the message for feasibility is clear:
unless the device is engineered to reduce effective curvature noise (material choice, screening, symmetric operation, and/or dynamical decoupling enabled by fast optical control), decoherence is likely to be the dominant constraint for gate fidelity.

Finally, it is useful to express a minimal ``coherent-control'' requirement in terms of the Rabi frequency.
From Eqs.~\eqref{eq:Heff} and~\eqref{eq:Rabi_scaling}, a resonant pulse drives oscillations at $\Omega_R=2\mathcal{E}_0|M_{3\to 5}^{(\alpha)}|/\hbar$ (up to envelope factors).
A necessary condition for high-contrast Rabi oscillations is therefore
\begin{equation}
\Omega_R\,T_2^*\;\gg\;1,
\label{eq:coherent_control_condition}
\end{equation}
while two-qubit gates require $J\,T_2^*\gg 1$.
Equations~\eqref{eq:J_estimate}--\eqref{eq:coherent_control_condition} provide a compact quantitative checklist: \emph{fast optical control is available in principle}, but turning the correlation-sector qubit into a high-fidelity quantum-computing primitive hinges on suppressing (or refocusing) the remaining quadrupolar charge-noise channels.

\section{Roadmap: from $N=2$ gate primitives to topological FQHE control}

The $N=2$ system is simultaneously (i)~a fully analytic gate primitive and (ii)~a stepping stone toward genuinely topological FQHE degrees of freedom at $N\ge 3$.

\subsection{What carries over robustly to larger $N$}

Three qualitative features are expected to generalize:

\begin{enumerate}
\item \textbf{OAM selection rules are symmetry-protected.}
The angular-momentum transfer $s=l+\sigma$ implies $\Delta L=\pm s$ for the many-body state, providing a tunable ladder between total-angular-momentum sectors.
\item \textbf{Kohn protection remains the baseline.}
Spatially uniform dipole probes ($l=0$) excite only CM motion, remaining blind to correlations in a parabolic dot.
Twisted light is therefore the minimal optical route to correlation spectroscopy.
\item \textbf{Line strengths are correlation diagnostics.}
The intensity of a TL-driven transition depends on interaction-defined eigenstates (pseudopotentials), so spectroscopy directly probes correlation content.\cite{Rodriguez2025}
\end{enumerate}

\subsection{Quantitative $N=3$ Calogero scaffold: $1/N$ expansion and the magic-number ladder}

For $N=3$ the pairwise inverse-square interaction
\begin{equation}
V_{\mathrm{Cal}}^{(N=3)}=\frac{\hbar^2\beta^2}{m^\ast}\sum_{i<j}\frac{1}{r_{ij}^2},
\label{eq:VCal_N3}
\end{equation}
admits a semi-analytic spectrum in a magnetic field via the resummed $1/N$ expansion of Ref.~\cite{GQR1996} (see also Ref.~\cite{Johnson1995}).
The center-of-mass still decouples exactly in a parabolic dot, so the relevant spectrum organizes by the \emph{internal} total angular momentum.
To avoid confusion with the optical OAM index $l$, we denote the many-body angular-momentum sector by $L\ge 0$ (standard FQHE convention); for $B>0$ this corresponds to the low-energy left-chiral branch of the dot spectrum.

In the notation of Ref.~\cite{GQR1996}, the internal energies can be written in the compact form
\begin{equation}
\begin{split}
E_{L;\,n_u,n_\triangle,m_\triangle} \approx \hbar\omega \biggl[ &\sqrt{L^2+6\beta^2} + \delta(L;n_\triangle,m_\triangle) \\
&+ 2n_u + 1 \biggr] - \frac{\hbar\omega_c}{2}L,
\end{split}
\label{eq:N3_energy_1overN}
\end{equation}
where $\omega=\sqrt{\omega_0^2+\omega_c^2/4}$, $n_u$ counts the symmetric ``breathing'' mode, and $(n_\triangle,m_\triangle)$ label an effective 2D ``shape'' mode of the triangle.
The shift $\delta(L;n_\triangle,m_\triangle)$ is an ${\cal O}(1)$ quantity fixed by a simple transcendental equation (given explicitly in Ref.~\cite{GQR1996}) and is easily obtained numerically.
Ref.~\cite{GQR1996} also provides explicit approximate eigenfunctions in the normal-mode coordinates, so TL matrix elements reduce to standard oscillator integrals.
Equation~\eqref{eq:N3_energy_1overN} is therefore a \emph{design-level} scaffold for $N=3$: it gives transition frequencies, anharmonicities, and parameter trends without full diagonalization.

\paragraph{Magic-number ladder and the $L=3k$ sequence.}
For three \emph{spin-polarized} fermions, Ref.~\cite{GQR1996} finds that the ground state progresses through the ``magic'' angular momenta
\begin{equation}
L_{\mathrm{gs}}(B)\in\{3,6,9,12,\dots\},
\label{eq:L3k}
\end{equation}
as the magnetic field is increased, reproducing the well-known incompressible/magic-number phenomenology of three-electron dots and agreeing qualitatively with Coulomb calculations.\cite{HawrylakPfannkuche1993}
In particular, the $q=3$ Laughlin precursor sits at $L=9$, and the adjacent $L=12$ manifold is precisely where the single-quasihole and the $\Delta L=3$ edge excitations live.

\paragraph{Why an $s=3$ twisted-light pulse is special at $N=3$.}
A quasihole at the origin is created by
\begin{equation}
\hat{Q}_{\mathrm{qh}}=\prod_{i=1}^3 z_i \equiv e_3(z_1,z_2,z_3),
\label{eq:qh_e3}
\end{equation}
which raises total angular momentum by $\Delta L=+3$.
A twisted-light mode with $s=l+\sigma=3$ therefore targets the correct sector \emph{directly} ($L=9\rightarrow 12$).
In the LLL-projected language, the simplest $s=3$ optical multipole is the power-sum symmetric polynomial
\begin{equation}
p_3(z)\equiv \sum_{i=1}^3 z_i^3 .
\label{eq:p3}
\end{equation}
Newton's identity for three variables,
\begin{equation}
p_3=e_1^3-3e_1e_2+3e_3,
\label{eq:newton_p3}
\end{equation}
makes the physics transparent: an $s=3$ drive necessarily contains a \emph{direct quasihole component} ($\propto e_3$) plus components in the other $\Delta L=3$ basis states ($e_1^3$ and $e_1e_2$).
Interactions select the eigenbasis within the $L\to L+3$ manifold, so the experimentally relevant question becomes quantitative: what is the overlap of the TL-driven superposition with the quasihole eigenstate, and can it be enhanced by spectral selectivity and pulse shaping?

\subsection{Realistic interactions vs.\ Calogero: what must be validated quantitatively}
\label{sec:calogero_vs_coulomb}

The Calogero model is powerful because it yields exact closed-form energies and matrix elements, but a quantum-computing proposal ultimately lives or dies on quantitative device-level parameters: the qubit splitting $\omega_{\mathrm{qubit}}$, the Rabi rates $\Omega_R$ and leakage rates set by the anharmonicity $\mathcal{A}$, and the two-qubit coupling $J$.
For generic Coulomb or screened interactions, the \emph{qualitative} chiral fingerprints and selection rules can persist\cite{Rodriguez2025}, but high-fidelity pulses require that the key quantities be either (i) close enough to Calogero predictions, or (ii) experimentally calibratable.

A practical viewpoint is that the Calogero analytics should be used as an \emph{initial design model} and as a way to identify which transitions are symmetry-allowed/forbidden and which sectors are targeted by a given $(l,\sigma)$.
Then, for a given device with a realistic interaction, one performs a short spectroscopic calibration using the same TL control channel:
the transition frequencies directly determine $\omega_{\mathrm{qubit}}$ and $\mathcal{A}$, while Rabi oscillations determine $\Omega_R$ for the chosen pulse shape.
This is completely standard in other qubit modalities (superconducting circuits, trapped ions), and it reduces the burden on the Calogero-to-Coulomb \emph{quantitative} match.

To make this explicit, Table~\ref{tab:Calogero_vs_Coulomb} lists the minimal set of quantities that must be known (or measured) to run gates, together with the closed-form Calogero expressions already derived in this paper and the corresponding ``how to obtain'' route for a Coulomb/screened dot.
For $N=2$, the Coulomb/screened problem is still one-dimensional in the relative radius $r$ for each $m$, so all entries can be produced numerically by straightforward radial diagonalization (or by exact diagonalization in a truncated Fock--Darwin basis) without the exponential complexity of many-body FQHE.

\begin{table*}[t]
\caption{Key gate-level quantities and how to obtain them in the analytic Calogero model vs.\ a realistic Coulomb/screened dot. The point is not that Calogero must be quantitatively exact, but that all required parameters are either closed-form (Calogero) or directly calibratable (device).}
\begin{center}
\renewcommand{\arraystretch}{1.15}
\begin{tabular}{p{0.18\textwidth}p{0.36\textwidth}p{0.36\textwidth}}
\hline\hline
Quantity & Calogero ($1/r^2$) & Coulomb/screened ($1/r$) \\ \hline
$\omega_{\mathrm{qubit}}$ & Eq.~\eqref{eq:qubit_freq} & TL spectroscopy (line positions) \\
$\mathcal{A}$ & Eq.~\eqref{eq:anharmonicity} & ED / radial solve + spectroscopy \\
$M_{3\to 5}$ & Eq.~\eqref{eq:matrix_element} & ED / radial solve (wavefunctions) \\
$\Omega_R$ & Eqs.~\eqref{eq:Heff} and~\eqref{eq:Rabi_scaling} & Rabi calibration (pulse area) \\
$\langle r^2\rangle_m$ & Eq.~\eqref{eq:r2_expectation} & ED / radial solve \\
$J$ & Eq.~\eqref{eq:J_estimate} & multipole + ED / calibration \\ \hline\hline
\end{tabular}
\end{center}
\label{tab:Calogero_vs_Coulomb}
\end{table*}

The remaining quantitative question is \emph{fidelity transfer}: how close must the Calogero-designed pulse be to the Coulomb-optimized pulse to achieve high fidelity?
Because our protocol is intrinsically spectroscopic, the most conservative answer is: it need not be close at all, provided the device is stable enough that $\omega_{\mathrm{qubit}}$ and $\Omega_R$ can be calibrated and remain stationary over the experiment.
The more optimistic possibility---to be tested numerically for representative device geometries---is that the relevant quantities (especially the selection-rule structure and relative strengths of the dominant matrix elements) vary smoothly across repulsive central interactions, so that Calogero provides not just qualitative guidance but a quantitatively good first guess.

\subsection{How the Gambardella \texorpdfstring{$\mathrm{SU}(1,1)$}{SU(1,1)} machinery can be leveraged for this project}

A central practical obstacle in scaling beyond the analytically solvable $N=2$ case is to retain an operator-level understanding of \emph{which} many-body sectors are bright/dark under twisted light and how the relevant transition frequencies and anharmonicities scale with interaction strength.
A powerful (and underused in the quantum-dot literature) route is to exploit the fact that inverse-square interactions belong to a broad class of Hamiltonians whose \emph{dynamical group} is $\mathrm{SU}(1,1)$.

Gambardella showed that for $N$ interacting particles in any spatial dimension, after eliminating the center of mass, the internal Hamiltonian
with a quadratic (harmonic) term plus an additional translation-invariant potential that is homogeneous of degree $-2$
has $\mathrm{SU}(1,1)$ as a spectrum-generating algebra: the Casimir eigenvalue fixes the ladder spectrum, and a family of eigenfunctions can be generated algebraically by raising operators acting on the ground state.\cite{Gambardella1975}
This includes (as special cases) Calogero--Sutherland-type inverse-square forces and also certain three-body and higher-body homogeneous interactions.\cite{Gambardella1975}

\paragraph{Why this is directly relevant to parabolic quantum dots.}
Even though quantum dots are defined by a \emph{one-body} parabolic confinement $V_{\mathrm{dot}}=\frac{1}{2}m^\ast\omega_0^2\sum_i r_i^2$ rather than an explicit pair potential, the identity
\begin{equation}
\sum_{i=1}^N r_i^2
=
N R^2+\frac{1}{N}\sum_{i<j} r_{ij}^2,
\qquad
R\equiv \frac{1}{N}\sum_{i=1}^N r_i,
\label{eq:sumSquares_identity}
\end{equation}
implies that \emph{after} center-of-mass separation the internal confinement is proportional to $\sum_{i<j}r_{ij}^2$.
Thus, the internal problem in a parabolic dot fits naturally into the ``quadratic pair + homogeneous $-2$'' class treated in Ref.~\cite{Gambardella1975} once the CM degree of freedom is removed.
The magnetic field adds the standard Fock--Darwin $L_z$ term; it lifts chiral degeneracies but does not eliminate the internal ladder structure (Sec.~IV already illustrates this explicitly for $N=2$).

\paragraph{\texorpdfstring{$\mathrm{SU}(1,1)$}{SU(1,1)} generators and ladder structure.}
In Jacobi coordinates $\{X_a\}_{a=1}^{N-1}$ for the internal motion (with conjugate momenta $\{P_a\}$), define the quadratic ``size'' operator, dilation operator, and scale-invariant Hamiltonian,
\begin{equation}
\begin{split}
\hat{C} = \frac{m^\ast}{2}\sum_{a=1}^{N-1} X_a^2, &\qquad \hat{D} = \frac{1}{2}\sum_{a=1}^{N-1}\!\left(X_a\!\cdot\! P_a+P_a\!\cdot\! X_a\right), \\
\hat{H}_0 &= \frac{1}{2m^\ast}\sum_{a=1}^{N-1} P_a^2+V_{\mathrm{hom}}(\{X_a\}),
\end{split}
\label{eq:SU11_generators_def}
\end{equation}
where $V_{\mathrm{hom}}$ is translation invariant and homogeneous of degree $-2$ (e.g.\ Calogero $1/r_{ij}^2$).
Then (under the homogeneity condition) these operators close an $\mathfrak{su}(1,1)\cong\mathfrak{so}(2,1)$ Lie algebra,\cite{Gambardella1975}
\begin{equation}
\begin{split}
[\hat{C}, \hat{D}] = 2i\hbar\hat{C}, &\qquad [\hat{C}, \hat{H}_0] = i\hbar\hat{D}, \\
[\hat{H}_0, \hat{D}] &= -2i\hbar\hat{H}_0.
\end{split}
\label{eq:SU11_commutators}
\end{equation}
The trapped internal Hamiltonian is $\hat{H}_{\mathrm{int}}=\hat{H}_0+\omega^2 \hat{C}$ (with an appropriate effective $\omega$ set by $\omega_0$ and $B$).
Standard linear combinations of $(\hat{H}_0,\hat{C},\hat{D})$ define $\mathrm{SU}(1,1)$ ladder operators $\hat{K}_\pm$ and a Cartan generator $\hat{K}_0$ such that $\hat{K}_+$ raises the ``breathing'' quantum number by one step.\cite{Gambardella1975}
Consequently, within each $\mathrm{SU}(1,1)$ tower the spectrum is a near-harmonic ladder with step $\simeq 2\hbar\omega$ (exact in the solvable class), while the tower label is set by the Casimir (equivalently, by the interaction-determined ``angular'' sector on the $(N\!-\!1)$-dimensional shape sphere).

\paragraph{How this helps twisted-light control design.}
For the purposes of this proposal, the Gambardella viewpoint supplies three concrete project tools:

\begin{enumerate}
\item \textbf{State construction beyond brute force.}
Ref.~\cite{Gambardella1975} gives a constructive scheme in which a family of excited states is generated by an energy-raising operator acting on the ground state.
This mirrors our $N=2$ Laguerre--tower structure (Sec.~IV) and provides an analytic scaffold for $N\ge 3$ breathing-mode ladders.

\item \textbf{Selection-rule bookkeeping.}
Twisted-light operators are low-order polynomials in the coordinates (LLL-projected multipoles).
Decomposing them into irreducible pieces with respect to $\mathrm{U}(1)$ rotations ($\Delta L=\pm s$) and the $\mathrm{SU}(1,1)$ ladder structure predicts, \emph{a priori}, which internal towers are addressable at given $s$ and which matrix elements are symmetry-forbidden.

\item \textbf{A bridge to realistic Coulomb dots.}
Even when the true Coulomb interaction breaks exact solvability, $\mathrm{SU}(1,1)$ remains a useful organizing principle for near-harmonic internal ladders.
The program is then to (i)~use the $\mathrm{SU}(1,1)$ scaffold to design pulses and expected spectral patterns, and (ii)~calibrate the remaining interaction-specific mixing by exact diagonalization in the LLL basis.
\end{enumerate}

\paragraph{Worked example: $N=3$ and the $s=3$ TL operator selects a \emph{finite set} of SU(1,1) towers.}
To make the ``tower selectivity'' concrete, consider the lowest-Landau-level polynomial language for $N=3$.
The Laughlin state at filling $\nu=1/3$ has polynomial $\mathcal{V}^3=\prod_{i<j}(z_i-z_j)^3$ and total angular momentum $L=3N(N-1)/2=9$.
A TL pulse with $s=l+\sigma=3$ (e.g.\ $l=2$, $\sigma=+1$) injects $\Delta L=+3$.
In the LLL-projected long-wavelength limit, the corresponding internal driving operator reduces to the symmetric power-sum polynomial
\begin{equation}
\hat{O}_{s=3}^{\mathrm{LLL}}\;\propto\;\sum_{j=1}^3 z_j^3 \;\equiv\; p_3(z_1,z_2,z_3),
\end{equation}
which maps the $L=9$ Laughlin state into the $L=12$ manifold.
For $N=3$ this manifold is three-dimensional, and a natural basis is given by multiplying $\mathcal{V}^3$ by the degree-3 symmetric polynomials
$\{e_3,\;e_1e_2,\;e_1^3\}$.
Using Newton's identity (Eq.~\eqref{eq:newton_p3}),
\begin{equation}
p_3 = e_1^3 - 3 e_1 e_2 + 3 e_3,
\end{equation}
we immediately see that a \emph{single} $s=3$ TL pulse does not create a pure quasihole state $e_3\mathcal{V}^3$; rather, it creates a specific superposition across \emph{three} candidate towers.

This is exactly where the SU(1,1) language earns its keep.
After separating the center of mass, $e_1$ is proportional to the CM coordinate ($e_1=3Z_{\mathrm{CM}}$), so the $e_1^3$ and $e_1e_2$ pieces contain CM excitation content, while $e_3$ is the quasihole factor.
In Gambardella's terms, each linearly independent homogeneous polynomial that solves the ``angular'' problem is a lowest-weight vector of some SU(1,1) representation; the internal breathing ladder is generated by $K_+$ acting on it.
Equation above therefore \emph{decomposes the TL drive into reduced matrix elements} connecting the Laughlin tower to a \emph{finite, explicitly identified set} of target towers.
All other towers in the same $L=12$ sector (if present at larger $N$) are ``dark'' in the idealized limit because their lowest-weight vectors are orthogonal to the polynomial content of $p_3$.
Interactions and non-idealities can mix these towers, but the polynomial decomposition provides a controlled starting point for which modes are expected to be bright/dark and why.

In short, Gambardella's dynamical-group framework turns ``$N\ge 3$ is hard'' into a structured workflow:
identify the relevant $\mathrm{SU}(1,1)$ towers (breathing vs.\ shape), compute the twisted-light selection rules and dominant overlaps, and then layer Coulomb realism on top.

\subsection{What becomes genuinely new at $N\ge 3$}

At $N\ge 3$, Laughlin-like droplets support quasiholes with fractional charge and anyonic exchange phases.\cite{Laughlin1983,NayakReview}
Optically addressing the $\Delta L=N$ quasihole sector is therefore the first step from ``correlation spectroscopy'' toward preparation of topological excitations.
For $N=3$, the preceding subsection makes this concrete: an $s=3$ twisted-light drive targets the $L=9\to 12$ manifold where the single-quasihole and the $\Delta L=3$ edge excitations reside.

To connect explicitly to standard FQHE wavefunction language, recall that a quasihole at the origin is created by multiplying the many-body polynomial by $\prod_{i=1}^N z_i$, which increases total angular momentum by $N$ and inserts one quantum of vorticity (one unit of flux) through the droplet.
Hence, for $N=3$ the operator $\hat{Q}_{\mathrm{qh}}=\prod_{i=1}^3 z_i$ raises $L$ by $+3$, and a twisted-light mode with $s=3$ (e.g.\ $l=2,\sigma=+1$) is the minimal optical multipole that can populate that sector in a single pulse.
If the relevant matrix element is sizable \emph{and} the $L=12$ eigenbasis is spectroscopically resolvable, this becomes an optical ``write'' primitive for a quasihole \emph{sector} (with selectivity set by interaction-defined overlaps, cf.\ Sec.~VIII-B).

\paragraph{Abelian anyons as a phase resource.}
Laughlin quasiholes are \emph{Abelian} anyons: exchanging two quasiholes yields statistical angle $\theta=\pi/q$ (a full braid gives $2\theta=2\pi/q$) but (on simply connected geometries with fixed quasihole positions) no fusion-space degeneracy, so braiding alone is not universal.\cite{NayakReview,KitaevFault}
Nevertheless, such phases can provide robust, geometry-insensitive conditional phases when combined with non-topological control.
In this sense, Abelian anyons provide a \emph{protected phase resource} even when they do not supply a universal gate set by braiding alone.\cite{NayakReview}

\paragraph{How twisted light might participate in braiding protocols.}
If twisted light can (i)~create quasihole sectors and (ii)~be spatially steered by a spatial light modulator, a natural speculative direction is to use structured light not only as a spectroscopy tool but as a way to \emph{move} localized charge deficits (quasiholes) in time.
In macroscopic FQHE samples, quasihole motion is typically controlled by electrostatic gates or local potentials; an optical route would be faster (pulse timescales) and potentially more flexible in geometry.
A key open question for this program is whether sufficiently localized optical potentials can be engineered without heating the electronic system out of the LLL.

\paragraph{Long-term direction.}
The ``photonic control bus'' viewpoint motivates extending the same TL-selective spectroscopy/state-preparation ideas to engineered dot states whose excitations are non-Abelian.\cite{NayakReview,KitaevFault}

\paragraph{Non-Abelian targets.}
Ultimately, universal topological quantum computation requires non-Abelian anyons.\cite{NayakReview,KitaevFault}
While the present work focuses on Laughlin-like correlations and their Calogero limits, the broader program is clear: identify correlated dot states (possibly in higher Landau levels or in engineered interactions) whose excitations support non-Abelian statistics, and test whether twisted light can selectively prepare and interrogate those sectors.

\subsection{Experimental roadmap}

A concrete near-term experimental roadmap is:

\begin{enumerate}
\item \textbf{Spectroscopy in a single dot ($N=2$).} Verify TL-only internal transitions and extract $\Delta m=\pm 2$ selection rules.\cite{Rodriguez2025}
\item \textbf{Coherent control.} Demonstrate Rabi oscillations between $|m|=3$ and $|m|=5$ sectors and Ramsey interferometry for phase control.
\item \textbf{Two-dot conditional phase.} Engineer state-dependent Coulomb coupling and test a $ZZ$-type entangling phase gate.
\item \textbf{Scale-up.} Extend to $N=3$ dots and test $s=3$ pumping into the quasihole sector as an optical state-preparation primitive.
\end{enumerate}

Even if the long-term target is topological quantum computing, the $N=2$ stage can already provide a complete, analytically characterized, optically driven single-qubit gate module---a rare benchmark system for bridging correlated-electron physics and quantum information science.


\section{Representative experimental numbers and feasibility}

To make the proposal concrete, we collect representative parameter values from the few-electron quantum-dot and ultrafast/THz literature.
The goal is not to lock in a single device design, but to show that the \emph{energy scales, magnetic fields, temperatures, and pulse bandwidths} required to drive the twisted-light (TL) selection rules are already standard in nanostructure experiments.

\begin{table*}[t]
\caption{Representative numbers supporting TL control of correlated sectors in few-electron quantum dots.}
\label{tab:numbers}
\centering
\small
\renewcommand{\arraystretch}{1.15}
\begin{tabular}{@{}p{0.22\textwidth}p{0.28\textwidth}p{0.42\textwidth}@{}}
\toprule
Quantity & Typical value & Comment / source \\
\midrule
Dot confinement energy & $\hbar\omega_{0}\sim 3~\mathrm{meV}$ &
Harmonic estimate used to model GaAs dots with $d_{\mathrm{eff}}\approx 100~\mathrm{nm}$ \cite{Kouwenhoven2001}. \\
Charging energy & $E_{c}\sim 1$--$4~\mathrm{meV}$ &
Screened vs.\ unscreened estimates for $d_{\mathrm{eff}}\approx 100~\mathrm{nm}$ (and a commonly used intermediate value $E_c\simeq 2~\mathrm{meV}$ in model spectra) \cite{Kouwenhoven2001}. \\
Magnetic field & $B\lesssim 16~\mathrm{T}$ &
Field range routinely applied in few-electron dot spectroscopy \cite{Kouwenhoven2001}. \\
Cyclotron energy (GaAs) & $\hbar\omega_{c}\simeq (1.73~\mathrm{meV/T})\,B$ &
Set by the effective mass $m^{*}=0.067m_{e}$ \cite{Kouwenhoven2001}. \\
Zeeman scale (GaAs) & $|g|\mu_{B}B\simeq (0.025~\mathrm{meV/T})\,B$ &
With $g_{\mathrm{GaAs}}=-0.44$ \cite{Kouwenhoven2001}. \\
Base temperature & $T\sim 100$--$200~\mathrm{mK}$ &
Typical measurement temperatures in few-electron dot transport spectroscopy \cite{Kouwenhoven2001}. \\
THz-TDS waveform & pulse $\lesssim 1~\mathrm{ps}$, bandwidth around $1~\mathrm{THz}$ &
Single-cycle THz pulses used in time-domain spectroscopy (photoconductive or optical-rectification based) \cite{Koch2023}. \\
Strong-field THz example & $E_{\mathrm{pk}}=236$--$819~\mathrm{kV/cm}$ &
High-repetition-rate DSTMS optical-rectification source (1--10~kHz) \cite{Zheng2024}. \\
THz vortex-beam components & $0.3$--$0.45~\mathrm{THz}$ ($|L|=1,2$);\ $0.62$--$1.3~\mathrm{THz}$ ($|L|=1$) &
Survey of demonstrated THz vortex shapers and mode converters (including reports of $|L|\le 3$ at $1.8$--$2.8~\mathrm{THz}$) \cite{Petrov2022}. \\
\bottomrule
\end{tabular}
\end{table*}

A useful conversion is $1~\mathrm{meV}/h \simeq 0.242~\mathrm{THz}$, so a typical confinement $\hbar\omega_{0}\sim 3~\mathrm{meV}$ corresponds to a natural scale $f_{0}\sim 0.73~\mathrm{THz}$.
In the quantum Hall (Fock--Darwin) regime, the low-energy ladder spacing between successive lowest-Landau-level orbitals is
\begin{equation}
\hbar\omega_{-}=\hbar\sqrt{\omega_{0}^{2}+\omegac^{2}/4}-\frac{\hbar\omegac}{2},
\end{equation}
so for $\hbar\omega_{0}=3~\mathrm{meV}$ and $B=3~\mathrm{T}$ (i.e.\ $\hbar\omegac\simeq 5.2~\mathrm{meV}$) one finds $\hbar\omega_{-}\approx 1.36~\mathrm{meV}$ ($f_{-}\approx 0.33~\mathrm{THz}$), while at $B=5~\mathrm{T}$ one finds $\hbar\omega_{-}\approx 0.93~\mathrm{meV}$ ($f_{-}\approx 0.22~\mathrm{THz}$), all within standard far-infrared/THz bandwidths \cite{Koch2023,Zheng2024}.
These are precisely the frequencies relevant for the $|m|\to |m|+s$ internal ladders that TL addresses.

For spatial scales, $\hbar\omega_{0}=3~\mathrm{meV}$ and $m^{*}\approx 0.067m_{e}$ imply a single-particle Fock--Darwin length $\ellzero=\sqrt{\hbar/(2m^{*}\omega_{0})}\approx 14~\mathrm{nm}$ at $B=0$. In a magnetic field the relevant Gaussian length uses $\omega=\sqrt{\omega_0^2+\omega_c^2/4}$, giving $\ellzero(B=5~\mathrm{T})\approx 10~\mathrm{nm}$, comparable to the magnetic length $\ellB=\sqrt{\hbar/(eB)}\approx 11.5~\mathrm{nm}$ at $B=5~\mathrm{T}$.
Hence the relevant internal-coordinate matrix elements are set by tens-of-nanometre length scales, consistent with few-electron dots in the $d_{\mathrm{eff}}\sim 100~\mathrm{nm}$ class \cite{Kouwenhoven2001}.

Finally, the optics side is compatible with the selection rules: the protocol requires only modest OAM charges ($l=1,2$ in most of the worked examples), and THz vortex converters with $|L|=1$ (and in some cases $|L|=2$) have been demonstrated in the $0.3$--$1.3~\mathrm{THz}$ range, with reports of $|L|\le 3$ at a few THz \cite{Petrov2022}.

\subsection{How the Gonz\'alez--Quiroga--Rodr\'iguez $N=3$ theory can scaffold the topological roadmap}
\label{sec:GQR_scaffold}

A main bottleneck in moving from the fully analytic $N=2$ case to genuinely topological droplets ($N\ge 3$) is that the correlated spectrum becomes multi-dimensional and brute-force exact diagonalization scales rapidly with Hilbert-space size.
A particularly useful intermediate step already exists in the few-body literature: Gonz\'alez, Quiroga, and Rodr\'iguez developed a resummed $1/N$ expansion (with $N$ essentially set by the internal angular momentum scale) for three particles in a 2D parabolic dot under magnetic field with a Calogero ($1/r^2$) repulsion.\cite{GQR1996}
For our purposes, the value is practical: their method yields a design-level semi-analytic spectrum \emph{and} explicit normal-mode wavefunctions (breathing vs.\ shape) that reduce twisted-light matrix elements to oscillator integrals, while reproducing the correct ``magic-number'' ground-state ladder and its magnetic-field-driven transitions in good qualitative agreement with Coulomb benchmarks.\cite{GQR1996,HawrylakPfannkuche1993}

\section{Conclusion}

Twisted light provides an optical handle on correlation physics in magnetized quantum dots by accessing relative-motion excitations that are dark to uniform dipolar probes.
In the analytically solvable Calogero limit, both the spectrum and the twisted-light matrix elements are closed-form functions of the interaction strength, allowing gate parameters to be written down explicitly.
This suggests that the $N=2$ system can provide a concrete single-qubit gate primitive (with a plausible two-qubit extension via state-dependent Coulomb coupling) and a controlled stepping stone toward $N\ge 3$ FQHE droplets where quasiholes become genuinely topological excitations.
The central message for quantum computing audiences is operational: twisted light enables \emph{WRITE} (pulse-create a correlation sector), \emph{READ} (spectroscopically diagnose correlations), and \emph{SCALE} (optical addressing via SLM) in a unified photonic control layer.

\section{Data availability}

This work is theoretical and analytical. The figures present the output from prior published numerical code that appeared in Ref. \cite{Rodriguez2025} and was made available from the authors on reasonable request. All equations, derivations, and parameter estimates are presented in full within the manuscript. The Calogero-model eigenstates and matrix elements (Eqs.~\eqref{eq:calogero_state}--\eqref{eq:matrix_element}) are closed-form analytic results that can be reproduced from the formulas given. The $N=3$ semi-analytic spectrum (Eq.~\eqref{eq:N3_energy_1overN}) is based on the $1/N$ expansion method of Ref.~\cite{GQR1996}, whose transcendental equation is fully specified therein. No experimental datasets were generated or analyzed. Representative parameter values (Table~\ref{tab:numbers}) are drawn from published sources cited in the text.

\section{Author contributions}

F.J.R.\ and L.Q.\  developed the twisted-light coupling calculations formalism and exact selection rules for quantum dots under a magnetic field. L.Q.\  and N.F.J.\ contributed to the Calogero-model analysis, the $N=3$ semi-analytic framework, the coupled-dot entanglement analysis and the quantum-computing mapping. All authors discussed and verified the results, wrote the manuscript and contributed to revisions.

\begin{acknowledgments}
F.J.R.\ and L.Q.\   are grateful for financial support from project 5228 Banco de la Republica (2025, Colombia) and Facultad de Ciencias, project INV-2025-213-3449, Universidad de Los Andes (Colombia).
\end{acknowledgments}

\end{document}